\documentclass[a4article,11pt]{article}
\usepackage{jheppub} 
 \usepackage[T1]{fontenc}
 \usepackage{booktabs}
 \usepackage{xcolor} 
 \usepackage{xspace}
 \usepackage{dcolumn}
 \usepackage{hyperref}
 \usepackage{caption}
 \usepackage
[subrefformat=parens,position=top,skip=-15pt,margin=15pt,justification=justified,singlelinecheck=false]
 {subcaption}
 \usepackage{placeins}   
 \usepackage[group-digits=false]{siunitx}
 \usepackage{microtype}
 \usepackage{lmodern}
 \usepackage{hyperref}

\pdfoutput=1

\allowdisplaybreaks[1]

\newcommand{\Pl}{\ell}

\newcommand{\fb}{{\ensuremath\unskip\,\text{fb}}\xspace}
\newcommand{\ab}{{\ensuremath\unskip\,\text{ab}}\xspace}
\def\refeq#1{\mbox{(\ref{#1})}}
\def\reffi#1{\mbox{Figure~\ref{#1}}}
\def\reffis#1{\mbox{Figures~\ref{#1}}}
\def\refta#1{\mbox{Table~\ref{#1}}}

\def\refse#1{\mbox{Section~\ref{#1}}}

\def\citere#1{\mbox{Ref.~\cite{#1}}}
\def\citeres#1{\mbox{Refs.~\cite{#1}}}

\newcommand{\ie}{\emph{i.e.}\ }

\def\be{\begin{equation}}
\def\ee{\end{equation}}

\newcommand{\PH}{\ensuremath{\text{H}}\xspace}
\newcommand{\Pj}{\ensuremath{\text{j}}\xspace}
\newcommand{\Pp}{\ensuremath{\text{p}}\xspace}
\newcommand{\Pe}{\ensuremath{\text{e}}\xspace}
\newcommand{\Pep}{\ensuremath{\text{e}^+}\xspace}
\newcommand{\Pem}{\ensuremath{\text{e}^-}\xspace}

\newcommand{\Pq}{\ensuremath{\text{q}}\xspace}
\newcommand{\Pt}{\ensuremath{\text{t}}\xspace}
\newcommand{\Pu}{\ensuremath{\text{u}}\xspace}
\newcommand{\Pd}{\ensuremath{\text{d}}\xspace}
\newcommand{\Ps}{\ensuremath{\text{s}}\xspace}
\newcommand{\Pc}{\ensuremath{\text{c}}\xspace}
\newcommand{\Pg}{\ensuremath{\text{g}}\xspace}

\newcommand{\PW}{\ensuremath{\text{W}}\xspace}
\newcommand{\PZ}{\ensuremath{\text{Z}}}

\newcommand{\Mt}{\ensuremath{m_\Pt}\xspace}

\newcommand{\MWOS}{\ensuremath{M_\PW^\text{OS}}\xspace}
\newcommand{\MW}{\ensuremath{M_\PW}\xspace}
\newcommand{\MZOS}{\ensuremath{M_\PZ^\text{OS}}\xspace}
\newcommand{\MZ}{\ensuremath{M_\PZ}\xspace}

\newcommand{\Gt}{\ensuremath{\Gamma_\Pt}\xspace}
\newcommand{\GH}{\ensuremath{\Gamma_\PH}\xspace}

\newcommand{\GZOS}{\ensuremath{\Gamma_\PZ^\text{OS}}\xspace}

\newcommand{\GWOS}{\ensuremath{\Gamma_\PW^\text{OS}}\xspace}

\newcommand{\GeV}{\ensuremath{\,\text{GeV}}\xspace}
\newcommand{\TeV}{\ensuremath{\,\text{TeV}}\xspace}

\newcommand{\sw}{s_{\mathrm{w}}}
\newcommand{\alphas}{\ensuremath{\alpha_\text{s}}\xspace}
\newcommand{\order}[1]{\ensuremath{\mathcal{O}{\left(#1\right)}}\xspace}

\newcommand{\GF}{\ensuremath{G_\mu}}

\newcommand{\ptsub}[1]{\ensuremath{p_{\text{T},#1}}\xspace}

\newcommand{\MVOS}{\ensuremath{M_{V}^\text{OS}}\xspace}%
\newcommand{\GVOS}{\ensuremath{\Gamma_{V}^\text{OS}}\xspace}%

\newcommand{\newc}{\newcommand}
\newc{\bi}{\begin{itemize}}
\newc{\ei}{\end{itemize}}
\newc{\benu}{\begin{enumerate}}
\newc{\eenu}{\end{enumerate}}
\newc{\bc}{\begin{center}}
\newc{\ec}{\end{center}}
\newc{\bfig}{\begin{figure}}
\newc{\efig}{\end{figure}}
\newc{\qbar}{\bar{q}}
\newc{\go}{\tilde{g}}
\newc{\PB}{\textsc{Powheg-Box}}

\newcommand{\recola}{{\sc Recola}\xspace}

\newcommand{\openloops}{O\protect\scalebox{0.8}{PEN}L\protect\scalebox{0.8}{OOPS}\xspace}

\newcommand{\mocanlo}{{\sc MoCaNLO}\xspace}
\newcommand{\mocanlorecola}{{\sc MoCaNLO+Recola}\xspace}
\newcommand{\collier}{{\sc Collier}\xspace}

\newcommand{\rT}{{\mathrm{T}}}
\newcolumntype{.}{D{.}{.}{-1}}
\newcolumntype{d}[1]{D{.}{.}{#1}}

\colorlet{tableoverheadcolor}{gray!37.5}
\colorlet{tableheadcolor}{gray!25}
\colorlet{tablerowcolor}{gray!12.5}

\newcommand{\gsim}
{\;\raisebox{-.3em}{$\stackrel{\displaystyle >}{\sim}$}\;}

\newlength{\width}
\newlength{\height}

\def\draftdate{\relax}
\def\mda{\relax}
\def\mua{\relax}
\def\mla{\relax}
\def\draft{
\def\thtystars{******************************}
\def\sixtystars{\thtystars\thtystars}
\typeout{}
\typeout{\sixtystars**}
\typeout{* Draft mode!
         For final version remove \protect\draft\space in source file *}
\typeout{\sixtystars**}
\typeout{}
\def\draftdate{\today}
\def\mua{\marginpar[\boldmath\hfil$\uparrow$]%
                   {\boldmath$\uparrow$\hfil}\color{black}%
                    \typeout{marginpar: $\uparrow$}\ignorespaces}
\def\mda{\color{red}\marginpar[\boldmath\hfil$\downarrow$]%
                   {\boldmath$\downarrow$\hfil}%
                    \typeout{marginpar: $\downarrow$}\ignorespaces}
\def\mla{\marginpar[\boldmath\hfil$\rightarrow$]%
                   {\boldmath$\leftarrow $\hfil}%
                    \typeout{marginpar: $\leftrightarrow$}\ignorespaces}
\def\Mua{\marginpar[\boldmath\hfil$\Uparrow$]%
                   {\boldmath$\Uparrow$\hfil}\color{black}%
                    \typeout{marginpar: $\uparrow$}\ignorespaces}
\def\Mda{\color{red}\marginpar[\boldmath\hfil$\Downarrow$]%
                   {\boldmath$\Downarrow$\hfil}%
                    \typeout{marginpar: $\downarrow$}\ignorespaces}
\def\Mla{\marginpar[\boldmath\hfil\textcolor{red}{$\Rightarrow$}]%
                   {\boldmath\textcolor{red}{$\Leftarrow $}\hfil}%
                    \typeout{marginpar: $\leftrightarrow$}\ignorespaces}
\overfullrule 5pt
\oddsidemargin 15mm
\marginparwidth 29mm
}

\title{\hfill ~\\[-58mm]
\phantom{h} \hfill\mbox{\small {Cavendish-HEP-20/11}, {VBSCAN-PUB-09-20}}
\\[1cm]
\vspace{13mm}   
NLO QCD and EW corrections to vector-boson scattering into ZZ at the LHC}

\subheader{\today}

\author{Ansgar Denner$^1$,}
\author{Robert Franken$^1$,}
\author{Mathieu Pellen$^2$,}
\author{Timo Schmidt$^1$}

\affiliation{$^1$Universit\"at W\"urzburg, %
        Institut f\"ur Theoretische Physik und Astrophysik, \\ %
        Emil-Hilb-Weg 22,  %
        97074 W\"urzburg, %
        Germany%
}
\affiliation{$^2$University of Cambridge, Cavendish Laboratory,
        Cambridge CB3 0HE, United Kingdom%
}
\emailAdd{ansgar.denner@physik.uni-wuerzburg.de}
\emailAdd{robert.franken@physik.uni-wuerzburg.de}
\emailAdd{mpellen@hep.phy.cam.ac.uk}
\emailAdd{timo.schmidt@physik.uni-wuerzburg.de}

\abstract{We present the first calculation of the full
  next-to-leading-order electroweak and QCD corrections for
  vector-boson scattering (VBS) into a pair of Z bosons at the LHC. We
  consider specifically the process
  $\Pp\Pp\to\Pe^{+}\Pe^{-}\mu^{+}\mu^{-}\Pj\Pj+X$ at orders
  $\order{\alpha^7}$ and $\order{\alpha_s\alpha^6}$ and take all
  off-shell and interference contributions into account. Owing to the
  presence of enhanced Sudakov logarithms, the electroweak corrections
  amount to $-16\%$ of the leading-order electroweak fiducial cross
  section and induce significant shape distortions of differential
  distributions. 
  The QCD corrections on the other hand are larger ($+24\%$) than typical QCD corrections in VBS.
  This originates from considering the full computation including
  tri-boson contributions in a rather inclusive phase space. 
  We also provide a leading-order analysis of all
  contributions to the cross section for
  $\Pp\Pp\to\Pe^{+}\Pe^{-}\mu^{+}\mu^{-}\Pj\Pj+X$ in a realistic
  setup.
  }

\begin{document}

\maketitle

\newpage

\section{Introduction}
The non-observation of physics not described by the Standard Model (SM) at the Large Hadron Collider (LHC)
shifts into the focus of interest tests of the SM of particle physics and in
particular of the mechanism of electroweak (EW) symmetry breaking. 
A particularly important class of
processes in this respect is vector-boson scattering (VBS). It is
not only sensitive to the scalar sector of the EW theory but
also to non-standard triple and quartic gauge-boson couplings.

The scattering of like-sign W bosons as well as WZ has been observed
\cite{Aad:2014zda, Khachatryan:2014sta, Aaboud:2016ffv,
  Sirunyan:2017ret,Aaboud:2018ddq,Sirunyan:2019ksz,Aaboud:2019nmv,
  Sirunyan:2020gyx} by both ATLAS and CMS in leptonic final states.
Quite recently, measurements of VBS into a pair of leptonically
decaying Z~bosons, which allows for a cleaner experimental detection
albeit with lower hadronic cross section, have been published by both
experiments \cite{Sirunyan:2017fvv, Aad:2020zbq,CMS:2020zly}.
Moreover, EW di-boson production in association with a high-mass
di-jet system in semileptonic final states has been searched for
\cite{Sirunyan:2019der,Aad:2019xxo}.

Theoretically, VBS has been studied since a long
time. NLO QCD corrections to all VBS processes exist for more than ten
years (see \citeres{Baglio:2014uba,Rauch:2016pai} and references
therein). On the other hand, EW corrections to these
processes have become available only recently
\cite{Biedermann:2016yds, Biedermann:2017bss, Denner:2019tmn}. Thereby
it has been found that EW corrections to fiducial cross sections of
VBS are generically at the level of $-15\%$ \cite{Biedermann:2016yds}.
This has been confirmed by complete calculations for the scattering of
$\PW^+\PW^+$ \cite{Biedermann:2017bss} and $\PW\PZ$ pairs
\cite{Denner:2019tmn}.  Furthermore, for like-sign W scattering an event generator
including EW and QCD corrections is available \cite{Chiesa:2019ulk}.

In this article, we present for the first time NLO EW corrections to
VBS into a pair of Z~bosons at the LHC, considering specifically the
fully leptonic final state $\Pe^{+}\Pe^{-}\mu^{+}\mu^{-}\Pj\Pj$. The
prospects of this channel for the high-luminosity and high-energy
upgrade of the LHC have been studied in \citere{CMS:2018mbt}.  NLO QCD
corrections to the purely EW VBS process in the so-called VBS
approximation, which neglects $s$-channel diagrams and interferences
between $u$-channel and $t$-channel diagrams, have been presented in
\citere{Jager:2006cp}. These corrections have been matched
\cite{Jager:2013iza} to a QCD parton shower via the {\tt POWHEG BOX}
framework \cite{Alioli:2010xd}.  NLO QCD corrections to the
corresponding QCD-induced process have been provided in
\citere{Campanario:2014ioa}. Very recently, loop-induced $\PZ\PZ$
production with up to 2 jets merged and matched to parton showers has
been studied \cite{Li:2020nmi}.

The leading-order (LO) cross section for
$\Pp\Pp\to\Pe^{+}\Pe^{-}\mu^{+}\mu^{-}\Pj\Pj+X$ receives contributions
of orders $\order{\alpha^6}$ (EW contributions),
$\order{\alpha_s\alpha^5}$ (interference), and
$\order{\alpha_s^2\alpha^4}$ (QCD-induced contributions). The VBS
subprocess is part of the gauge-invariant EW contributions.  In the
experimental analysis, the three contributions are often referred to
as signal, interference, and QCD background, respectively.  While the
QCD-induced contributions are larger than the EW
ones for typical VBS cuts, the interference is smaller than the
latter. Since gluon-induced contributions at the order
$\order{\alpha_s^4\alpha^4}$ contribute sizeably, we include them in
our analysis.  In this paper we focus on the NLO EW and QCD
corrections to the LO EW contributions. More precisely, we consider
the complete gauge-invariant set of contributions at orders
$\order{\alpha^7}$ and $\order{\alpha_s\alpha^6}$. While the former
are pure EW corrections to the LO EW contributions, the latter receive
contributions from the QCD corrections to the LO EW contributions as
well as from EW corrections to the LO interferences.  In our NLO
calculation we do not rely on approximations but take into account the
full set of diagrams relevant at the corresponding perturbative order
including all interference contributions.  We also compare the full
NLO EW corrections with the ones obtained within a Sudakov
approximation.

The final state $\Pe^{+}\Pe^{-}\mu^{+}\mu^{-}\Pj\Pj$
leads to a wide-spread variety of contributing partonic channels.
This renders the full calculation at NLO EW and especially at
NLO QCD technically very demanding and CPU-time intensive.

This article is structured as follows: In \refse{sec:process} the
considered process and the different contributions are described. In
addition, details of our calculation including the checks
to validate our results are presented.  Section \ref{sec:results}
contains numerical results and their description. Lastly, in
\refse{sec:conclusion} a summary and concluding remarks are given.

\section{Description of the calculation}
\label{sec:process}

\subsection{Leading-order contributions}
\label{ssec:locontributions}

We are studying the process
\begin{equation}
\label{eq:LO-process}
\Pp\Pp\rightarrow{\Pe^{+}\Pe^{-}\mu^{+}\mu^{-}\Pj\Pj}+X.
\end{equation}
At LO, VBS appears in quark-induced partonic channels $\mathrm{q
  q}\rightarrow\mathrm{e^{+}e^{-}\mu^{+}\mu^{-} q q}$ (q generically
stands for a quark or anti-quark). The amplitudes for these processes
receive contributions of order $\order{g^6}$ as well as of order
$\order{g_\mathrm{s}^2g^4}$, where $g$ and $g_\mathrm{s}$ denote
the EW and strong coupling constant, respectively.

We neglect quark mixing and use a unit quark-mixing matrix. Since a
non-trivial quark-mixing matrix would only affect the $s$-channel
contributions and the NLO corrections, its effects are suppressed.

\begin{figure}
\setlength{\parskip}{1ex}
\begin{subfigure}[t]{0.33\textwidth}
\centering
\captionsetup{skip=0pt}
\caption{}
\includegraphics[page=1,scale=0.9]{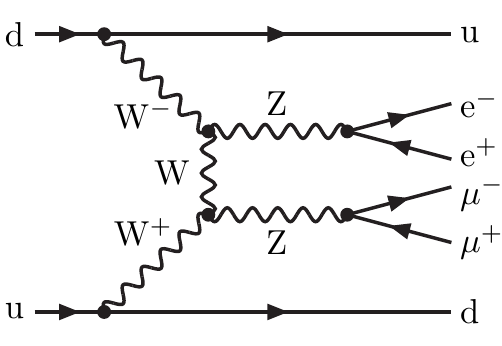}
\label{fig:born_qq_vbs}
\end{subfigure}
\begin{subfigure}[t]{0.33\textwidth}
\centering
\captionsetup{skip=0pt}
\caption{}
\includegraphics[page=2,scale=0.9]{Diagrams/diagrams.pdf}
\label{fig:born_qq_Higgs}
\end{subfigure}%
\begin{subfigure}[t]{0.33\textwidth}
\centering
\captionsetup{skip=0pt}
\caption{}
\includegraphics[page=3,scale=0.9]{Diagrams/diagrams.pdf}
\label{fig:born_qq_doubres}
\end{subfigure}%
\par
\begin{subfigure}[c]{0.33\textwidth}
\centering
\captionsetup{skip=0pt}
\caption{}
\includegraphics[page=4,scale=0.9]{Diagrams/diagrams.pdf}
\label{fig:born_qq_singlres}
\end{subfigure}%
\begin{subfigure}[c]{0.33\textwidth}
\centering
\captionsetup{skip=0pt}
\caption{}
\includegraphics[page=5,scale=0.9]{Diagrams/diagrams.pdf}
\label{fig:born_qq_nonres}
\end{subfigure}%
\begin{subfigure}[c]{0.33\textwidth}
\centering
\captionsetup{skip=0pt}
\caption{}
\includegraphics[page=6,scale=0.9]{Diagrams/diagrams.pdf}
\label{fig:born_s_channel_W}
\end{subfigure}%
\par
\begin{subfigure}{0.33\textwidth}
\centering
\captionsetup{skip=0pt}
\caption{}
\includegraphics[page=7,scale=0.9]{Diagrams/diagrams.pdf}
\label{fig:born_qq_triW}
\end{subfigure}
\begin{subfigure}{0.33\textwidth}
\centering
\captionsetup{skip=0pt}
\caption{}
\includegraphics[page=8,scale=0.9]{Diagrams/diagrams.pdf}
\label{fig:born_qq_triZga}
\end{subfigure}%
\begin{subfigure}{0.33\textwidth}
\centering
\captionsetup{skip=0pt}
\caption{}
\includegraphics[page=9,scale=0.9]{Diagrams/diagrams.pdf}
\label{fig:born_QCD}
\end{subfigure}%
\caption{Examples of tree-level Feynman diagrams.}
\label{fig:borndiagrams}
\end{figure}

Some sample diagrams of order $\order{g^6}$ are shown in
\reffis{fig:born_qq_vbs}--\ref{fig:born_qq_triZga}.
The diagrams in \reffis{fig:born_qq_vbs} and \ref{fig:born_qq_Higgs}
represent the characteristic $t$-channel VBS topology. While the first
diagram illustrates the scattering of $\PW^+\PW^-$ into $\PZ\PZ$, the
second one exemplifies $\PZ\PZ$ scattering, which takes place only via
Higgs exchange.
Higgs-exchange $s$-channel diagrams arise in VBS into ZZ as a new feature with
respect to same-sign W and WZ scattering.%
\footnote{For our set of cuts, the Higgs resonance does not appear in
  the fiducial phase-space region.}
  However, partonic channels
involving only $\PZ\PZ$ scattering are suppressed with respect to
those involving the subprocess $\PW^+\PW^-\to\PZ\PZ$.
Doubly-resonant, singly-resonant, and non-resonant
diagrams of non-VBS type contribute to the partonic processes at order
$\order{g^6}$, as illustrated in \reffis{fig:born_qq_doubres},
\ref{fig:born_qq_singlres}, and \ref{fig:born_qq_nonres}, respectively.
Besides $t$- and $u$-channel diagrams also $s$-channel diagrams show up
for some partonic processes (\reffi{fig:born_s_channel_W}).
In particular, $s$-channel diagrams corresponding to triple gauge-boson
production, both for $\PW\PZ\PZ$ and $\PZ\PZ\PZ$ production appear as
depicted in \reffis{fig:born_qq_triW} and \ref{fig:born_qq_triZga}.
Lastly, diagrams at the order $\order{g^4g_s^2}$ are
characterised by $t$-channel gluon exchange between the two quark
lines (see \reffi{fig:born_QCD}).
While in the diagrams we specify only \PZ~bosons in
  $s$-channel propagators to stress the contributions to the
  intermediate $\PZ\PZ$ state, corresponding diagrams with photons are
  also included in our computation. In fact, we take into account the complete set of 
  diagrams for the given six-fermion final state.

Consequently, at the level of squared amplitudes three kinds of
gauge-invariant contributions exist, purely EW contributions at
$\order{\alpha^6}$, QCD-induced contributions at order
$\order{\alphas^2\alpha^4}$ and contributions at
order $\order{\alphas\alpha^5}$, which result from the
interference of diagrams at order $\order{g^6}$ and
$\order{g_\mathrm{s}^2g^4}$.
Owing to the colour structure, the latter contributions are only
non-vanishing, if diagrams of two different kinematic channels ($s$,
$t$, $u$) with all 4 external quarks in the same generation are
interfered, such as the diagrams in \reffi{fig:born_s_channel_W} and
\reffi{fig:born_QCD}.

We do not consider contributions with bottom quarks in the initial
state, which are PDF suppressed, and also do not include final states
with bottom quarks.%
\footnote{We verified that the contributions of bottom
quarks are below 3\% for our inclusive setup.}
Then 60 partonic quark-induced channels contribute compared to 40 for
WZ and 12 for $\PW^\pm\PW^\pm$ scattering (not counting $\Pq\Pq'$ and
$\Pq'\Pq$ initial states separately).  Out of these 60 channels, 24 
receive non-vanishing interference contributions between
different coupling orders that make up the contribution
of order $\order{\alphas\alpha^5}$.  At order
$\order{\alphas^2\alpha^4}$, in addition to the 60 quark-induced
channels, channels with one or two gluons in the initial state
contribute.  Given the large gluon luminosity at the LHC, the latter
are one of the reasons for the enhancement of the QCD-induced
contributions over the EW ones.

Further contributions at orders $\order{\alpha^6}$ and
$\order{\alphas\alpha^5}$ result from photon-induced processes with
$\gamma\gamma$, $\gamma\Pg$ and $\gamma\Pq$ initial states. Such
contributions were found to be below $0.5\%$ for $\PW\PZ$ scattering
\cite{Denner:2019tmn}, which is also expected for VBS into $\PZ\PZ$.
These contributions are neglected in this work.

\begin{figure}
\centering
\begin{subfigure}[b]{0.33\textwidth}
\centering
\captionsetup{skip=0pt}\caption{}
\includegraphics[page=22,scale=0.9]{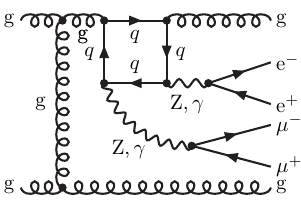}
\label{fig:loop_induced_4_point_function} 
\end{subfigure}%
\begin{subfigure}[b]{0.33\textwidth}
\centering
\captionsetup{skip=0pt}\caption{}
\includegraphics[page=23,scale=0.9]{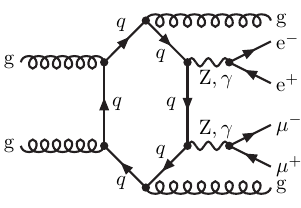}
\label{fig:loop_induced_6_point_function}
\end{subfigure}%
\caption{Sample diagrams for the loop-induced process 
$\Pg\Pg\to\Pe^{+}\Pe^{-}\mu^{+}\mu^{-}\Pg\Pg$.}
\label{fig:loop_induced}
\end{figure}

In contrast to final states corresponding to charged $\PW^\pm\PW^\pm$ and
$\PW\PZ$ scattering, the $\Pe^{+}\Pe^{-}\mu^{+}\mu^{-}\Pj\Pj$ final
state receives contributions from the loop-induced partonic
process $\Pg\Pg\to\Pe^{+}\Pe^{-}\mu^{+}\mu^{-}\Pg\Pg$ at order
$\order{\alphas^4\alpha^4}$ (see \reffi{fig:loop_induced} for sample
diagrams).  We include these contributions in our leading-order
analysis.

\subsection{Virtual corrections}
\label{ssec:virtualcorrections}
We compute NLO corrections  of orders $\order{\alpha^7}$ and $\order{\alphas\alpha^6}$
to the process \refeq{eq:LO-process}.
\begin{figure}
\begin{subfigure}[b]{0.33\textwidth}
\centering
\captionsetup{skip=0pt}\caption{}
\includegraphics[page=10,scale=0.9]{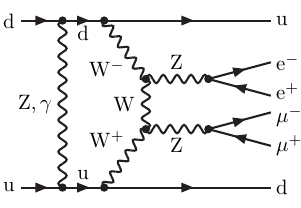}
\label{fig:virt_6point}
\end{subfigure}%
\begin{subfigure}[b]{0.33\textwidth}
\centering
\captionsetup{skip=0pt}\caption{}
\includegraphics[page=11,scale=0.9]{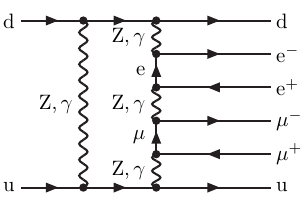}
\label{fig:virt_8point}
\end{subfigure}
\begin{subfigure}[b]{0.33\textwidth}
\centering
\captionsetup{skip=0pt}\caption{}
\includegraphics[page=12,scale=0.9]{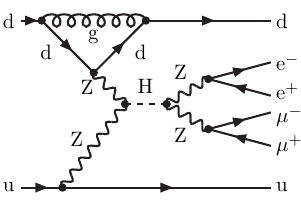}
\label{fig:virt_3point}
\end{subfigure}%
\par
\begin{subfigure}[b]{0.33\textwidth}
\centering
\captionsetup{skip=0pt}\caption{}
\includegraphics[page=13,scale=0.9]{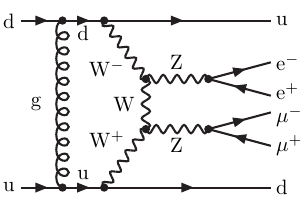}
\label{fig:virt_6point_gluon}
\end{subfigure}%
\begin{subfigure}[b]{0.33\textwidth}
\centering
\captionsetup{skip=0pt}\caption{}
\includegraphics[page=14,scale=0.9]{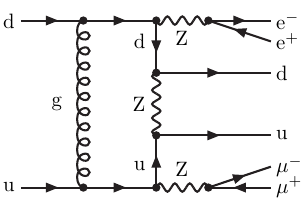}
\label{fig:virt_6point_doubly_res}
\end{subfigure}%
\begin{subfigure}[b]{0.33\textwidth}
\centering
\captionsetup{skip=0pt}\caption{}
\includegraphics[page=15,scale=0.9]{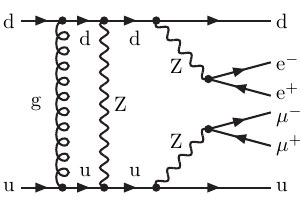}
\label{fig:virt_4point}
\end{subfigure}%
\caption{Sample one-loop diagrams.}
\label{fig:virtdiagrams}
\end{figure}

Virtual corrections of order $\order{\alpha^7}$ result from
interfering purely EW loop diagrams of order $ \order{g^8}$ with
tree diagrams of order $\order{g^6}$ and are thus EW corrections
to the LO EW diagrams. Examples for loop diagrams of order 
$\order{g^8}$ are depicted 
\reffis{fig:virt_6point}--\ref{fig:virt_8point}. While corrections to VBS
involve diagrams with up to 6-point functions (\reffi{fig:virt_6point}),
non-resonant diagrams contain up to 8-point functions
(\reffi{fig:virt_8point}).

Virtual corrections of order $\order{\alphas\alpha^6}$ have different
sources. First, loop diagrams of order $\order{g^6g_\mathrm{s}^2}$
interfere with tree diagrams of order $\order{g^6}$. Such loop
diagrams (see \reffis{fig:virt_3point}--\ref{fig:virt_4point} for
examples) can be viewed as QCD corrections to diagrams of order
$\order{g^6}$.  However, the diagrams in
\reffis{fig:virt_6point_doubly_res}--\ref{fig:virt_4point} can also be
viewed as EW corrections to diagrams of order
$\order{g^4g_\mathrm{s}^2}$.  As a consequence EW and QCD-induced
contributions cannot be separated at order
$\order{\alphas\alpha^6}$ on the basis of Feynman diagrams.  While diagrams with gluons
attached to a single quark line (see \reffi{fig:virt_3point})
contribute for all partonic channels, diagrams with gluon exchange
between different quark lines only contribute for partonic processes
that receive contributions of different kinematic channels such as those
in \reffis{fig:virt_6point_doubly_res} and \ref{fig:virt_4point},
which interfere with the LO diagram \reffi{fig:born_qq_vbs}.
In the VBS approximation only $t$- and $u$-channel diagrams are taken
into account and no interferences between different kinematic
channels.  Then all corrections of $\order{\alphas\alpha^6}$ can
be interpreted as QCD corrections to the EW LO diagrams.

We do not include contributions of partonic channels with external
bottom quarks or photons in the initial state in the virtual corrections.

\subsection{Real corrections}
\label{ssec:realcorrections}

\begin{figure}
\begin{subfigure}[b]{0.33\textwidth}
\centering
\captionsetup{skip=0pt}\caption{}
\includegraphics[page=16,scale=0.9]{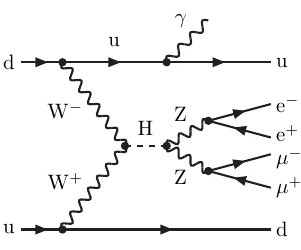}
\label{fig:Higgs real-photon radiation} 
\end{subfigure}
\begin{subfigure}[b]{0.33\textwidth}
\centering
\captionsetup{skip=0pt}\caption{}
\includegraphics[page=17,scale=0.9]{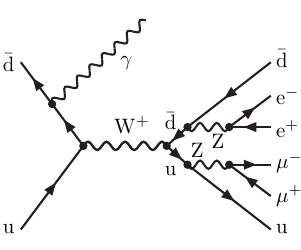}
\label{fig:W_s_channel_real_photon}
\end{subfigure}%
\begin{subfigure}[b]{0.33\textwidth}
\centering
\captionsetup{skip=0pt}\caption{}
\includegraphics[page=18,scale=0.9]{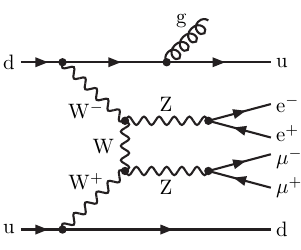}
\label{fig:real_gluon}
\end{subfigure}%
\par
\begin{subfigure}[b]{0.33\textwidth}
\centering
\captionsetup{skip=0pt}\caption{}
\includegraphics[page=19,scale=0.9]{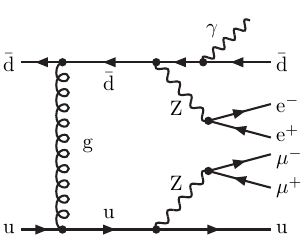}
\label{fig:QCD_real_photon}
\end{subfigure}%
\begin{subfigure}[b]{0.33\textwidth}
\centering
\captionsetup{skip=0pt}\caption{}
\includegraphics[page=20,scale=0.9]{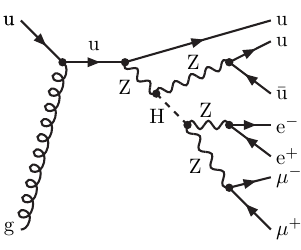}
\label{fig:gluon_ind_1}
\end{subfigure}%
\begin{subfigure}[b]{0.33\textwidth}
\centering
\captionsetup{skip=0pt}\caption{}
\includegraphics[page=21,scale=0.9]{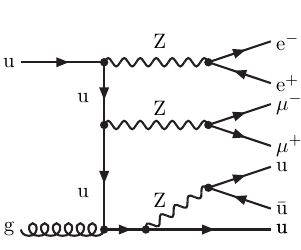}
\label{fig:gluon_ind_2}
\end{subfigure}%
\caption{Sample diagrams for real corrections.}
\label{fig:realdiagrams}
\end{figure}

In addition to the virtual corrections also real photon and gluon
emission needs to be considered. Some related Feynman diagrams are
shown in \reffi{fig:realdiagrams}.

In the quark-induced channels, emission of a real photon from a quark,
a W~boson, or a charged lepton gives rise to the partonic processes
$\Pq\Pq\rightarrow{\Pe^{+}\Pe^{-}\mu^{+}\mu^{-}\Pq\Pq\gamma}$.  These
furnish the real EW corrections to the LO EW diagrams with diagrams of
order $\order{g^7}$.  As examples, the diagrams in
\reffis{fig:Higgs real-photon radiation} and
\ref{fig:W_s_channel_real_photon} represent two of many possibilities
of a photon insertion into an $s$-channel Higgs-exchange and
$\PW$-boson exchange diagram, respectively. Squaring the sum of all
these diagrams yields the real corrections of order
$\order{\alpha^7}$.

As the virtual corrections, also the real contributions at order
$\order{\alphas\alpha^6}$ emerge from different sources.
First, as part of the NLO QCD corrections, emission of a real gluon
from quarks of the LO EW diagrams exemplarily depicted in
\reffi{fig:real_gluon}, results in diagrams of order
$\order{g^6g_\mathrm{s}}$.  These contribute to the partonic
process $\Pq\Pq\rightarrow{\Pe^{+}\Pe^{-}\mu^{+}\mu^{-}\Pq\Pq\Pg}$ and
upon squaring yield contributions of order
$\order{\alphas\alpha^6}$.  Contributions of the same order
result from interferences of real-photon-emission diagrams of orders
$\order{g^7}$ (\reffi{fig:W_s_channel_real_photon}) and
$\order{g^5g_\mathrm{s}^2}$ (\reffi{fig:QCD_real_photon})
corresponding to different kinematic channels.
Besides quark-induced channels also $\Pg\Pq$ channels appear with
diagrams resulting from crossing the gluon and a quark in
$\Pq\Pq\rightarrow{\Pe^{+}\Pe^{-}\mu^{+}\mu^{-}\Pq\Pq\Pg}$, as for
example shown in Figures~\ref{fig:gluon_ind_1} and
\ref{fig:gluon_ind_2}, and contribute at order $\order{\alphas\alpha^6}$.

The number of partonic channels for the quark-induced processes is the
same as for the corresponding LO processes. For the gluon-induced
$\Pg\Pq$ processes 40 partonic channels contribute compared to 28 for
WZ and none for $\PW^\pm\PW^\pm$ scattering. Note that we do not take
into account partonic channels with external bottom quarks.

Crossing the photon and a quark in
$\Pq\Pq\rightarrow{\Pe^{+}\Pe^{-}\mu^{+}\mu^{-}\Pq\Pq\gamma}$ gives
rise to photon-induced partonic processes that contribute both at orders
$\order{\alpha^7}$ and $\order{\alphas\alpha^6}$. Such
contributions have been found to be below $2\%$ for same-sign W~scattering
\cite{Biedermann:2017bss} and are neglected in this calculation.

For the treatment of the infrared (IR) singularities arising in the
phase-space integration of the real corrections we use Catani--Seymour
dipole subtraction for QCD \cite{Catani:1996vz} and its variant for
QED \cite{Dittmaier:1999mb,Dittmaier:2008md}.  In this regard, the
strategy for $\PW\PZ$ scattering of \citere{Denner:2019tmn} can be
taken over, and no new features appear.  At the order
$\order{\alphas\alpha^6}$ both QCD and QED IR singularities due to
soft and/or collinear gluon or photon emission or via forward
branchings of QCD partons in the initial state appear.  Another
type of singularity arises in diagrams with photons converting into a
quark--anti-quark pair in the final state.  The low virtuality of the
photon leads to a collinear singularity, which would cancel against
the virtual corrections to the $\Pe^{+}\Pe^{-}\mu^{+}\mu^{-}\Pj\gamma$
final state. While this final state is considered separately from VBS,
this singularity can be absorbed into a photon-to-jet conversion
function \cite{Denner:2019zfp}, which can be related to the
non-perturbative hadronic vacuum polarisation.  In general, different
kinds of singularities appear in the same diagrams and have to be
dealt with simultaneously. More details on the treatment of IR
singularities can be found in
\citeres{Biedermann:2017bss,Denner:2019tmn}.

\subsection{Details of the computation and validation}
\label{sec:details}

The results presented here have been obtained from the combination
\mocanlorecola of two computer programs.  \mocanlo is a generic Monte
Carlo integration code that can compute arbitrary processes at NLO QCD
and EW accuracy in the SM.  The use of phase-space mappings similar to
the ones of \citeres{Berends:1994pv,Denner:1999gp,Dittmaier:2002ap}
ensures an efficient integration even for high-multiplicity processes.
\recola~\cite{Actis:2012qn,Actis:2016mpe}, on the other hand, is a
tree and one-loop matrix-element provider.  It uses the \collier
library~\cite{Denner:2014gla,Denner:2016kdg} to obtain numerically the
one-loop scalar
\cite{'tHooft:1978xw,Beenakker:1988jr,Dittmaier:2003bc,Denner:2010tr}
and tensor integrals
\cite{Passarino:1978jh,Denner:2002ii,Denner:2005nn}.  The tandem
\mocanlorecola has already been used for many complex computations, in
particular for VBS processes
\cite{Biedermann:2016yds,Biedermann:2017bss,Ballestrero:2018anz,Denner:2019tmn,Pellen:2019ywl}.
It has also been successfully compared for WZ scattering with results
of {\sc BONSAY+}\openloops \cite{Denner:2019tmn} and for same-sign W
scattering with the combination of \recola and the Monte Carlo
integration code {\sc BBMC} \cite{Biedermann:2017bss}.  \mocanlorecola
has also served as baseline for the validation of the implementation
of $\order{\alpha^7}$ corrections in {\sc Powheg+}\recola
\cite{Chiesa:2019ulk} for same-sign W scattering and of
$\order{\alphas\alpha^5}$ corrections for $\Pp\Pp\to\PW\PW\Pj$ in {\sc
  Sherpa+}\recola \cite{Biedermann:2017yoi,Brauer:2020kfv}.  For VBS
into \PZ\PZ, we compared  the dominant partonic
process $\Pu\Pd\to\Pe^{+}\Pe^{-}\mu^{+}\mu^{-}\Pu\Pd$ as well as the
channels $\Pu\Ps\to\Pe^{+}\Pe^{-}\mu^{+}\mu^{-}\Pd\Pc$,
$\Pu\Pu\to\Pe^{+}\Pe^{-}\mu^{+}\mu^{-}\Pu\Pu$, and
$\Pu\bar\Pc\to\Pe^{+}\Pe^{-}\mu^{+}\mu^{-}\Pd\bar\Ps$ at order
$\mathcal{O}{\left(\alpha^7\right)}$ and selected contributions at
order $\mathcal{O}{\left(\alphas\alpha^6\right)}$ between
\mocanlorecola and {\sc BBMC+}\recola and found agreement within
integration errors.

In \mocanlo, the IR divergences in the real radiation are subtracted
using the dipole method for QCD \cite{Catani:1996vz} and its extension
to QED \cite{Dittmaier:1999mb,Dittmaier:2008md}.  Within the
dipole-subtraction scheme, the $\alpha_{\mathrm{dipole}}$
parameter~\cite{Nagy:1998bb} allows to restrict the phase space to the
singular regions.  The value $\alpha_{\mathrm{dipole}}=1$ corresponds
to the full phase space (within the acceptance defined by selection
cuts) without additional restrictions.  In the present case, two full
computations both at order $\order{\alpha^7}$ and $\order{\alphas
  \alpha^6}$ have been performed, one with
$\alpha_{\mathrm{dipole}}=10^{-2}$ and one with
$\alpha_{\mathrm{dipole}}=1$.  Both computations agree within
statistical errors at each order and provide thus a robust check of
the subtraction procedure.  The results presented here have been
obtained from the computation with $\alpha_{\mathrm{dipole}}=10^{-2}$.
In addition, representative contributions have also been computed with
two different numerical values of the IR-regulator parameter, proving
IR finiteness of the results.  Finally, for the computation of the
one-loop amplitudes of order $\order{g^6 g_{\rm s}^2}$, the two
different modes of {\sc Collier} have been used: the {\sc DD} mode in
the $\alpha_{\mathrm{dipole}}=1$ computation and the {\sc COLI} mode
in the $\alpha_{\mathrm{dipole}}=10^{-2}$ computation.  Throughout,
the massive resonant particles are described in the complex-mass
scheme \cite{Denner:1999gp,Denner:2005fg,Denner:2006ic} at both 
tree and one-loop level.

\section{Numerical Results}
\label{sec:results}

\subsection{Input parameters and event selection}

\subsubsection*{Input parameters}
\label{ssec:InputParameters}

The numerical simulations have been carried out for the LHC at a
centre-of-mass energy of $13\TeV$.  As parton distribution functions
(PDF) we employ the NLO NNPDF-3.1 Lux QED set with $\alphas(\MZ) =
0.118$ \cite{Ball:2014uwa,Bertone:2017bme}.  This is incorporated in
the calculation through
LHAPDF~\cite{Andersen:2014efa,Buckley:2014ana}.  We are working with
the fixed $N_\text{F}=5$ flavour scheme throughout.  Both the EW and
QCD collinear initial-state splittings are treated by ${\overline{\rm
    MS}}$ redefinition of the PDF.  Note that the same PDF set is used
for both the LO and NLO predictions.

The renormalisation and factorisation scales are set to the geometric
average of the transverse momentum of the jets
\begin{equation}
\label{eq:defscale}
 \mu_{\rm ren} = \mu_{\rm fac} = \sqrt{p_{\rm T, j_1}\, p_{\rm T, j_2}},
\end{equation}
where $\Pj_1$ and $\Pj_2$ are the two hardest identified jets (see
definition below) ordered according to transverse momentum.  In the
following, we perform the usual 7-point scale variation of both the
renormalisation and factorisation scale, \ie we calculate the
observables for the pairs of renormalisation and factorisation scales
\begin{equation}
(\mu_\mathrm{ren}/\mu_0,\mu_\mathrm{fact}/\mu_0) = (0.5,0.5),
(0.5,1),(1,0.5),(1,1),(1,2),(2,1),(2,2)
\end{equation}
with the central scale defined in \refeq{eq:defscale} and use the
resulting envelope.

To fix the electromagnetic coupling, the $G_\mu$ scheme
\cite{Denner:2000bj} is used, which defines the coupling from
the Fermi constant as
\begin{equation}
  \alpha = \frac{\sqrt{2}}{\pi} G_\mu \MW^2 \left( 1 - \frac{\MW^2}{\MZ^2} \right)  \qquad \text{with}  \qquad   {\GF    = 1.16638\times 10^{-5}\GeV^{-2}}.
\end{equation}
The masses and widths of the massive particles are chosen as
\cite{Tanabashi:2018oca}
\begin{alignat}{2}
\label{eqn:ParticleMassesAndWidths}
                  \Mt   &=  173.0\GeV,      & \quad \quad \quad \Gt &= 0 \GeV,  \nonumber \\
                \MZOS &=  91.1876\GeV,      & \quad \quad \quad \GZOS &= 2.4952\GeV,  \nonumber \\
                \MWOS &=  80.379\GeV,       & \GWOS &= 2.085\GeV,  \nonumber \\
                M_{\rm H} &=  125.0\GeV,    &  \GH  &=  4.07 \times 10^{-3}\GeV.
\end{alignat}
The bottom quark is taken to be massless, and no
partonic channel with initial-state and/or final-state bottom quarks
is included.  Since there are no resonant top quarks in the
considered processes, we set the top-quark width to zero.  The values
of Higgs-boson mass and width are taken from
\citere{Heinemeyer:2013tqa}.  The pole masses and widths of the W
and Z~bosons used in the calculation are determined from the measured
on-shell (OS) values \cite{Bardin:1988xt} via
\begin{equation}
        M_V = \frac{\MVOS}{\sqrt{1+(\GVOS/\MVOS)^2}}\,,\qquad  
\Gamma_V = \frac{\GVOS}{\sqrt{1+(\GVOS/\MVOS)^2}}.
\end{equation}

\subsubsection*{Event selection}

The event selection used here is inspired by the CMS measurement
\cite{Sirunyan:2017fvv}.  Experimentally, the final state of the
process is described by four charged leptons and at least two QCD
jets.  The QCD partons are combined into jets with the
anti-$k_\text{T}$ algorithm \cite{Cacciari:2008gp} using $R=0.4$ as
jet-resolution parameter.  In the same way, the real photons are
recombined with the final-state quarks into jets or with the charged
leptons into dressed leptons.  In both cases the anti-$k_\text{T}$
algorithm and a resolution parameter $R=0.4$ is utilised.

The four charged leptons $\ell$ are required to fulfil
\begin{align}
 \ptsub{\Pl} >  20\GeV,\qquad |\eta_{\Pl}| < 2.5, \qquad 
\Delta R_{\Pl\Pl'} > 0.05, \qquad M_{\Pl^+\Pl^{\prime-}} > 4 \GeV,
\end{align}
where $\Pl$ and $\Pl'$ can be any type of leptons and $\Pl^+$ and
$\Pl^{\prime-}$ have to be oppositely charged leptons regardless of
flavour.  In addition, a cut on the invariant mass of the leptonic
decay products of the Z bosons is applied
\begin{align}
\label{eq:mz}
  60 \GeV < M_{\Pl^+\Pl^-} < 120 \GeV,\qquad \Pl=\Pe,\mu.
\end{align}
Note that this cut also removes the phase-space region containing the
Higgs resonance.

After jet clustering, jets that fulfil the conditions
\begin{align}
 \ptsub{\Pj} >  30\GeV, \qquad |\eta_\Pj| < 4.7,\qquad\Delta R_{\Pj\Pl} > 0.4
\end{align}
are called identified jets, and at least two of them are required.
Note that the last condition demands a minimal distance between the
jet and all four charged leptons.  The two identified jets with
highest transverse momenta, called hardest or leading jets, should
obey
\begin{align}
\label{eq:vbscuts}
 M_{\Pj_1 \Pj_2} >  100\GeV .
\end{align}

\subsection{Cross sections}

In this section, numerical results are discussed for the fiducial
cross section in the setup defined above (inclusive setup for short)
as well as corresponding results with a stronger cut on the invariant
mass of the two leading jets (VBS setup)
\begin{align}
\label{eq:vbscuts2}
 M_{\Pj_1 \Pj_2} >  500\GeV .
\end{align}

We start by presenting contributions of different orders in the strong
and EW coupling constants at LO, \ie the EW component [order
$\order{\alpha^6}]$, the interference [order
$\order{\alphas\alpha^5}$], the QCD component [order
$\order{\alphas^2\alpha^4}$], and the loop-induced contributions
[order $\order{\alphas^4\alpha^4}$] in \refta{tab:LO}.
\begin{table}
\sisetup{group-digits=false}
\centering
\begin{tabular}{cccccc}
\toprule
Order
    & $\order{           \alpha^6 }$
    & $\order{ \alphas   \alpha^5 }$
    & $\order{ \alphas^2   \alpha^4 }$
    & $\order{ \alphas^4   \alpha^4 }$
    & Sum \\
\midrule
\multicolumn{2}{l}{$M_{\Pj_1\Pj_2}> 100\GeV$} \\
\midrule
$\sigma_{\mathrm{LO}} [\si{\femto\barn}]$
    & \num{ 0.097683 +- 0.000002}
    & \num{ 0.008628 +- 0.000001}
    & \num{ 1.062478 +- 0.000048}
    & \num{ 0.12101 +- 0.00064}
    & \num{ 1.28980 +- 0.00064}
    \\
$\textrm{fraction} [\si{\percent}]$
    & \num{7.57}
    & \num{0.67}
    & \num{82.38}
    & \num{9.38}
    & \num{100} \\
\midrule
\multicolumn{2}{l}{$M_{\Pj_1\Pj_2}> 500\GeV$} \\
\midrule
$\sigma_{\mathrm{LO}} [\si{\femto\barn}]$
    & \num{ 0.073676  +- 0.000003}
    & \num{ 0.005567  +- 0.000001}
    & \num{ 0.136143  +- 0.000015 }
    & \num{ 0.01345  +- 0.00029}
    & \num{ 0.22883  +- 0.00029 }
    \\
$\textrm{fraction} [\si{\percent}]$
    & \num{32.20}
    & \num{2.43}
    & \num{59.49}
    & \num{5.88}
    & \num{100} \\
\bottomrule
\end{tabular}
\caption{LO cross section (Sum) and contributions of individual
orders $\order{\alpha^6}$, $\order{\alphas\alpha^5}$,
$\order{\alphas^2\alpha^4}$, and $\order{ \alphas^4 \alpha^4 }$
for $\Pp\Pp \to \Pe^+\Pe^-\mu^+\mu^-\Pj\Pj+X$ at $13\TeV$ CM energy.
Photon-induced contributions and contributions with external bottom
quarks are not included.
Each contribution is given in $\fb$ and as fraction relative to the sum of the four contributions (in percent).
While the numbers in the upper part of the table are for the inclusive
setup, those in the lower part are for the VBS setup.
The digits in parentheses indicate integration errors.}
\label{tab:LO}
\end{table}
In the experimental analysis, the first three contributions are often
referred to as signal, interference, and QCD background, respectively.
We note that the contributions of bottom quarks, which are not included
in the results of \refta{tab:LO}, amount to $0.2\%$, $-1.1\%$ and
$2.8\%$ of the three contributions, respectively.
The most striking fact is the size of the QCD contributions with
respect to the EW ones.  In our default setup, the QCD component is an
order of magnitude larger, and the EW component containing VBS
contributions is only about $0.1\fb$.  This is in contrast to
same-sign-WW and WZ scattering which have larger cross sections as
well as larger signal-to-background ratios
\cite{Biedermann:2017bss,Denner:2019tmn}.  Imposing the stronger cut
$M_{\Pj_1\Pj_2}>500\GeV$, the QCD background is suppressed and is only
twice as large as the EW contribution, which is only moderately
decreased.\footnote{A similar effect could have been achieved
    by imposing a sensible cut on $\Delta y_{\Pj_1\Pj_2}$, as can be
    inferred from \reffi{fig:dyjj}.} The relevance of the
  interference grows from $1\%$ to 
$2.5\%$ of the cross section, while the relative contribution of the
loop-induced process decreases from $10\%$ to $6\%$.

Including 7-point scale variations as defined above, the EW
contribution to the LO cross section of order $\order{\alpha^6}$ reads
\begin{align}
 \sigma_{\alpha^6} ={} & 0.097683(2)^{+6.8\%}_{-6.0\%}
\qquad \text{ for } \quad M_{\Pj_1\Pj_2}> 100\GeV,\notag\\
 \sigma_{\alpha^6} ={} & 0.073676(3)^{+8.6\%}_{-7.5\%} 
\qquad \text{ for } \quad M_{\Pj_1\Pj_2}> 500\GeV.
\end{align}
  
In \refta{tab:NLO} we present results for the NLO cross section
including EW corrections of order $\order{\alpha^7}$, QCD corrections of order
$\order{\alphas\alpha^6}$, or both as well as the corresponding
corrections normalised to the LO cross section of order
$\order{\alpha^6}$ in percent.
\begin{table}
\sisetup{group-digits=false}
\centering
\begin{tabular}{cccc}
\toprule
Order
    & $\order{         \alpha^6 }+\order{         \alpha^7 }$
    & $\order{         \alpha^6 }+\order{ \alphas \alpha^6 }$
    & $\order{         \alpha^6 }+\order{         \alpha^7 } + \order{ \alphas \alpha^6 }$ \\
\midrule
\multicolumn{2}{l}{$M_{\Pj_1\Pj_2}> 100\GeV$} \\
\midrule
$\sigma_{\mathrm{NLO}} [\si{\femto\barn}]$
    & \num{0.08211 +- 0.00004 } 
    & \num{0.12078 +- 0.00011 } 
    & \num{0.10521 +- 0.00011} \\
$\sigma^{\mathrm{max}}_{\mathrm{NLO}} [\si{\femto\barn}]$
    & \num{0.08728 +- 0.00005 } $[+6.3\%]$
    & \num{0.12540 +- 0.00013 } $[+3.8\%]$
    & \num{0.10838 +- 0.00014 } $[+3.0\%]$\\
$\sigma^{\mathrm{min}}_{\mathrm{NLO}} [\si{\femto\barn}]$
    & \num{0.07749 +- 0.00004 } $[-5.6\%]$
    & \num{0.11656 +- 0.00009 } $[-3.5\%]$
    & \num{0.10225 +- 0.00009}  $[-2.8\%]$\\
$\delta [\si{\percent}]$
    & \num{-15.9}
    & \num{+23.6}
    & \num{+7.7} \\
\midrule
\multicolumn{2}{l}{$M_{\Pj_1\Pj_2}> 500\GeV$} \\
\midrule
$\sigma_{\mathrm{NLO}} [\si{\femto\barn}]$
    & \num{0.06069 +- 0.00004} 
    & \num{0.07375 +- 0.00025}  
    & \num{0.06077 +- 0.00025}  \\
$\sigma^{\mathrm{max}}_{\mathrm{NLO}} [\si{\femto\barn}]$
    & \num{0.06568 +- 0.00005 }  $[+8.2\%]$
    & \num{0.07466 +-0.00026}  $[+1.2\%]$
    & \num{0.06149 +- 0.00024 }  $[+1.2\%]$\\
$\sigma^{\mathrm{min}}_{\mathrm{NLO}} [\si{\femto\barn}]$
    & \num{0.05636 +- 0.00004}  $[-7.1\%]$
    & \num{0.07282+- 0.00021}  $[-1.3\%]$
    & \num{0.05977+- 0.00030}  $[-1.6\%]$\\
$\delta [\si{\percent}]$
    & \num{-17.6}
    & \num{0.1}
    & \num{-17.5} \\
\bottomrule
\end{tabular}
\caption{
Fiducial cross sections for $\Pp\Pp \to \Pep\Pem\mu^+\mu^-\Pj\Pj+X$ at
 $13\TeV$ CM energy at NLO EW  [$\order{\alpha^6
  }+\order{\alpha^7 }$], 
  NLO QCD [$\order{\alpha^6}
  + \order{\alphas \alpha^6 }$], 
  and NLO QCD+EW  [$\order{\alpha^6 }
  +\order{\alpha^7 } + \order{ \alphas \alpha^6 }$].
Each contribution is given in $\fb$ (with the extrema resulting from
scale variations as absolute numbers and as deviation in percent) and
as relative correction 
$\delta=\sigma_{\rm NLO} / \sigma_{\alpha^6}-1$ in percent.
While the numbers in the upper part of the table are for the inclusive
setup, those in the lower part are for the VBS setup.
The digits in parentheses indicate the integration errors.}
\label{tab:NLO}
\end{table}
We also list the numbers corresponding to the maxima and minima of the
7-point scale variation in absolute terms and relative to the central
values in percent. The scale dependence is reduced by a factor of 2 for
the inclusive setup and even more in the setup with the additional VBS
cut when including the QCD corrections of order
$\order{\alphas\alpha^6}$.  We find negative EW corrections at the
level of $16\%$--$17\%$, \ie in 
the same range as for other VBS processes \cite{Biedermann:2016yds,
  Denner:2019tmn}. While the relative EW corrections are only
marginally affected by the additional cut, the QCD corrections are
drastically reduced. The reason for this sizeable effect is discussed
further below.

We turn to the discussion of different partonic channels. To this end,
we split all partonic channels into 4 subsets: VBS-WW encompasses the
$16$ partonic channels that contain $\PW\PW\to\PZ\PZ$ as subprocess,
VBS-ZZ is made up of the remaining 32 channels that include the
$\PZ\PZ\to\PZ\PZ$ subprocess. The left-over channels are further
separated into 4 that contain $\Pp\Pp\to\PW\PZ\PZ$ as subprocess (WZZ)
and 8 that then always include the $\Pp\Pp\to\PZ\PZ\PZ$ subprocess
(ZZZ). We note that in total 36 partonic channels involve
$\PZ\PZ\to\PZ\PZ$, 8 involve $\PW\PZ\PZ$, and 16 involve $\PZ\PZ\PZ$.
None of the channels involves both $\PW\PW\to\PZ\PZ$ and $\PW\PZ\PZ$.

The contributions of these different partonic
processes are compiled in \refta{tab:subprocesses}, where we show the
corresponding contributions of orders $\order{\alpha^6}$,
$\order{\alpha^7 }$, and $\order{\alphas \alpha^6 }$ in $\mathrm{ab}$,
as well as the NLO corrections in percent.
\begin{table}
\sisetup{group-digits=false}
\centering
\begin{tabular}{cccccc}
\toprule
Contribution
    & $\sigma_{\alpha^6} [\si{\atto\barn}]$
    & $\Delta\sigma_{\alpha^7} [\si{\atto\barn}]$
    & $\Delta\sigma_{\alpha^7}/\sigma_{\alpha^6} [\si{\percent}]$
    & $\Delta\sigma_{\alphas\alpha^6} [\si{\atto\barn}]$
    & $\Delta\sigma_{\alphas\alpha^6}/\sigma_{\alpha^6} [\si{\percent}]$
\\\midrule
\multicolumn{2}{l}{$M_{\Pj_1\Pj_2}> 100\GeV$} \\
\midrule
all
    & \num{97.683 +- 0.002}
    & \num{-15.55 +- 0.05}  & \num{-15.9}
    & \num{23.10 +- 0.11}  & \num{23.6}
\\
VBS-WW 
    & \num{95.237 +- 0.002}
    & \num{-15.28 +- 0.05}  & \num{-16.0}
    & \num{1.33 +- 0.11}  & \num{1.4}
\\
VBS-ZZ
    & \num{1.9463 +- 0.0002}
    & \num{-0.1979 +- 0.0006} & \num{-10.2}
    & \num{3.892 +- 0.004} & \num{200}
\\  
WZZ
    & \num{0.1361 +- 0.0001}
    & \num{-0.0142 +- 0.0001} & \num{-10.5}
    & \num{13.850 +- 0.004} & \num{10174} 
\\
ZZZ
    & \num{0.3629 +- 0.0001}
    & \num{-0.0542 +- 0.0006} & \num{-14.9}
    & \num{4.029 +- 0.003} & \num{1110}
\\\midrule
\multicolumn{2}{l}{$M_{\Pj_1\Pj_2}> 500\GeV$} \\
\midrule
all
    & \num{73.679 +- 0.002}
    & \num{-13.01 +- 0.04}  & \num{-17.7}
    & \num{0.07 +- 0.25}  & \num{0.10}
\\
VBS-WW 
    & \num{72.846 +- 0.002}
    & \num{-12.91 +- 0.04}  & \num{-17.7}
    & \num{-2.73 +- 0.25}  & \num{-3.7}
\\
VBS-ZZ
    & \num{0.8096 +- 0.0002}
    & \num{-0.0986 +- 0.0003} & \num{-12.2}
    & \num{0.486 +- 0.006} & \num{60.1}
\\  
WZZ
    & \num{0.00471 +- 0.00002}
    & \num{-0.00085 +- 0.00001} & \num{-18.1}
    & \num{1.849 +- 0.005} & \num{39258} 
\\
ZZZ
    & \num{0.01887 +- 0.00001}
    & \num{-0.00529 +- 0.00002} & \num{-28.0}
    & \num{0.470 +- 0.001} & \num{2488}
\\\bottomrule
\end{tabular}
\caption{
Contributions of partonic channels to fiducial cross sections for
$\Pp\Pp \to \Pep\Pem\mu^+\mu^-\Pj\Pj+X$ at 
 $13\TeV$ CM energy. Contributions at orders $\order{\alpha^6}$, $\order{\alpha^7 }$, 
and $\order{\alphas \alpha^6 }$ are given  in $\ab$  and relative to
the corresponding $\order{\alpha^6}$ cross section in percent.
While the numbers in the upper part of the table are for the inclusive
setup, those in the lower part are for the VBS setup.
The digits in parentheses indicate the integration errors.}
\label{tab:subprocesses}
\end{table}
The LO $\order{\alpha^6}$ cross section is dominated by the $16$
partonic channels containing $\PW\PW\to\PZ\PZ$ as subprocess. The
remaining partonic channels contribute about $2.5\%$ and $1.0\%$ in
the inclusive and VBS setup, respectively, at LO and similarly at the
order $\order{\alpha^7}$. The relative EW corrections are smaller for
the non-VBS-WW channels than for the VBS-WW channels apart from ZZZ in
the VBS setup, which is however very small.  The
$\order{\alphas\alpha^6}$ contributions, on the other hand, are
dominated by channels involving triple-vector-boson production in the
inclusive setup. In the inclusive setup more than 70\% of the VBS-ZZ
contribution in the fifth column results from partonic channels that
also involve $\PW\PZ\PZ$.  Note that at this order also $\Pg\Pq$
channels contribute at the same level as the $\Pq\Pq$ channels and are
included in columns 5 and 6 of \refta{tab:subprocesses}. In the
VBS-setup, the VBS channels and the non-VBS channels practically
cancel at order $\order{\alphas\alpha^6}$. The cut
$M_{\Pj_1\Pj_2}>500\GeV$ reduces the $\order{\alphas\alpha^6}$
contributions of the WZZ/ZZZ channels by almost an order of magnitude. Note
that the QCD corrections are small for the dominating VBS-WW channels,
but huge for the WZZ/ZZZ channels.  The huge QCD corrections result
from contributions with three resonant vector bosons that are present
at NLO QCD but not at LO (see below).%
\footnote{The huge QCD corrections raise the question about
    the relevance of the non-trivial quark-mixing matrix. Out of the
    $24\%$ QCD corrections, $14\%$ result from partonic channels with
    (anti-)quark--gluon in the initial states. For these channels,
    owing to the unitarity of the CKM matrix, 
    the effects of a non-trivial quark-mixing matrix only result from
    mixing of the first two generations with the top quark, which is
    very small. The leading effect of quark-mixing results from the 
    quark--antiquark $s$-channel  contributions and is of order 
    $|V_{\Pu\Ps}/V_{\Pu\Pd}|^2\sim |V_{\Pc\Pd}/V_{\Pc\Ps}|^2\sim 5\%$. Since these
    channels cause $10\%$ QCD corrections, the effect of 
    neglecting quark mixing is at the level of $10\%\times5\%=0.5\%$.}

In \refta{tab:part_channel_EW} we show the relative EW corrections of
order $\order{\alpha^7}$ for selected partonic processes. 
\begin{table}
\sisetup{group-digits=false}
\centering
\begin{tabular}{cccccc}
\toprule
Part. channel 
    & $\sigma_{\mathrm{\alpha^6}} [\si{\atto\barn}]$
    & $\delta_{\mathrm{\alpha^7}}  [\si{\percent}]$
    & $\delta_{\mathrm{LL}}  [\si{\percent}]$
    & $\delta_{\mathrm{LL+SSC}} [\si{\percent}]$
    & subprocesses \\
\midrule
$\mathrm{u}\mathrm{d} \to \Pep\Pem\mu^+\mu^-\mathrm{u}\mathrm{d}$    
    & \num{ 51.537 +- 0.002 } 
    & \num{-17.3 +- .1}
    & \num{-16.4}
    & \num{-14.6} 
    & VBS-WW/VBS-ZZ\\
$\mathrm{u}\mathrm{s} \to 
\Pep\Pem\mu^+\mu^-\mathrm{d}\mathrm{c}$    
    & \num{ 12.769 +- 0.001 } 
    & \num{-15.1 +- .1}
    & \num{-14.2}
    & \num{-12.6} 
    & VBS-WW\\
$\mathrm{u}\mathrm{\bar{u}} \to
\Pep\Pem\mu^+\mu^-\mathrm{d}\mathrm{\bar{d}}$  
    & \num{ 10.666 +- 0.001} 
    & \num{-15.0 +- .1}
    & \num{-13.6}
    & \num{-10.1} 
    & VBS-WW/ZZZ\\
$\mathrm{u}\mathrm{u} \to \Pep\Pem\mu^+\mu^-\mathrm{u}\mathrm{u}$ 
    & \num{0.37718 +- 0.00005} 
    & \num{-11.8 +- .1}
    & --
    & --
    & VBS-ZZ\\
$\mathrm{u}\mathrm{\bar{d}} \to
\Pep\Pem\mu^+\mu^-\mathrm{u}\mathrm{\bar{d}}$  
    & \num{ 0.24011 +- 0.00005} 
    & \num{-10.2 +- .1}
    & --
    & -- 
    & WZZ\\
$\mathrm{u}\mathrm{\bar{u}} \to
\Pep\Pem\mu^+\mu^-\mathrm{u}\mathrm{\bar{u}}$   
    & \num{ 0.15878 +- 0.00004} 
    & \num{-11.6 +- .1}
    & --
    & -- 
    & VBS-ZZ/ZZZ\\
$\mathrm{d}\mathrm{\bar{d}} \to
\Pep\Pem\mu^+\mu^-\mathrm{s}\mathrm{\bar{s}}$  
    & \num{ 0.11638 +- 0.00003} 
    & \num{-11.0 +- .1}
    & --
    & -- 
    & ZZZ\\
\bottomrule
\end{tabular}
\caption{
Relative EW-NLO corrections
$\delta_{\alpha^7}=\Delta\sigma_{\alpha^7}/\sigma_{\alpha^6}$ for
various partonic 
channels and corresponding corrections in the Sudakov approximation
without ($\delta_{\mathrm{LL}}$) and with ($\delta_{\mathrm{LL+SSC}}$)
    angular-dependent logarithms.
    Contributing subprocesses are indicated in the last column, and
    digits in parentheses indicate the integration errors.
}
\label{tab:part_channel_EW}
\end{table}
In the LO cross sections $\sigma_{\mathrm{\alpha^6}}$ the
contributions of $\Pq\Pq'$ and $\Pq'\Pq$ are combined and the channels
resulting from interchanging all quarks of the first and second
generations are included. The last column shows the subprocesses
present in the channels. We have selected those channels with the
relevant subprocesses that give the largest contributions.  While the
relative EW corrections for the channels involving VBS-WW
are in the range $15\%$--$18\%$, they are between $7\%$ and $19\%$ for
the other channels.

The large EW corrections for the dominant VBS-WW channels can be
explained based on a Sudakov approximation applied to the
$\PW\PW\to\PZ\PZ$ subprocesses, as was already shown for like-sign
WW~scattering \cite{Biedermann:2016yds} and WZ scattering
\cite{Denner:2019tmn}. Following
\citeres{Denner:2000jv,Accomando:2006hq} one can show that the leading
logarithmic corrections to the scattering of transverse vector bosons,
which is the dominant contribution, yields the simple correction
factor
\begin{equation}
 \delta_{\textrm{LL}} = \frac{\alpha}{4\pi} \left\{
- 4 C_W^{\mathrm{EW}} \log^2 \left(\frac{Q^2}{\MW^2}\right)
+  2b_W^{\mathrm{EW}} \log \left(\frac{Q^2}{\MW^2}\right) \right\}.
\label{eq:LLcorr}
\end{equation}
It includes all logarithmically enhanced EW corrections apart from the
angular-dependent subleading soft-collinear logarithms and applies to
all VBS processes that are not mass suppressed, such as
$\PW\PW\to\PZ\PZ$, owing to the fact that these scattering processes
result from the same $\mathrm{SU}(2)_\mathrm{w}$ coupling.  Equation
\refeq{eq:LLcorr} is, however, not valid for mass-suppressed processes
like $\PZ\PZ\to\PZ\PZ$. The constants are given by $ C_W^{\mathrm{EW}}
= 2/\sw^2$ and $ b_W^{\mathrm{EW}} = 19/(6\sw^2)$, where $\sw$
represents the sine of the weak mixing angle.  Further, $Q$ is a
representative scale of the $VV \to VV$ scattering process, which can
conveniently be chosen as the four-lepton invariant mass $M_{4\ell}$.
Using $Q=M_{4\ell}$ event by event, results in the numbers for
$\delta_{\textrm{LL}}$ shown in the 4th column of
\refta{tab:part_channel_EW}, which agree within $2\%$ with the exact
NLO results. It should be clear that the formula \refeq{eq:LLcorr}
is not applicable to the non-VBS-WW processes.  For the angular-dependent
subleading soft-collinear logarithms, correction factors can be
derived as well in the Sudakov limit based on the results of
\citeres{Denner:2000jv,Accomando:2006hq}. These depend on the specific
VBS process, and the correction factor for $\PW\PW\to\PZ\PZ$ reads
\begin{equation}\label{eq:WWZZ_SSC_corr_TTTT}
 \delta_{\text{SSC}} = \frac{\alpha}{\pi\sw^2} 
 2\ln\left(\frac{Q^2}{\MW^2}\right)
\left[-\ln\frac{s_{12}}{Q^2}
+\frac{s_{23}}{s_{12}}\ln\frac{s_{13}}{Q^2}
+\frac{s_{13}}{s_{12}}\ln\frac{s_{23}}{Q^2}
\right],
\end{equation}
where $s_{12}$, $s_{13}$, and $s_{23}$ are the Mandelstam variables of
the VBS process.\footnote{To determine the Mandelstam variables, we
  need the momenta of the scattering vector bosons. The momentum of
  one of them is fixed by combining the momentum $p_1$ of one of the
  incoming partons with the momentum $p_{\mathrm{parton,out}}$  
  of the outgoing parton that
  maximises $(p_1-p_{\mathrm{parton,out}})^2$. The momentum of the
  second scattering vector boson is obtained from the momentum of the
  second incoming parton and the other outgoing parton.}  Applying the
correction factor \refeq{eq:WWZZ_SSC_corr_TTTT} combined with
\refeq{eq:LLcorr} event by event yields
the approximations for the EW corrections in column 5 of
\refta{tab:part_channel_EW}. While the inclusion of these
angular-dependent logarithmic terms somewhat deteriorates the
agreement with the full correction factors, it shows that the effect
of the angular-dependent logarithmic corrections is only at the level
of $2\%$ and thus well within the accuracy of the approximation, which
is expected to be a few percent.

\subsection{Differential distributions}

In this section we discuss distributions for the inclusive setup
\refeq{eq:vbscuts} only.

\subsubsection*{LO distributions}

\begin{figure}
\setlength{\parskip}{-4ex}
\begin{subfigure}{0.49\textwidth}
\centering
\subcaption{}
\includegraphics[width=1.\linewidth]{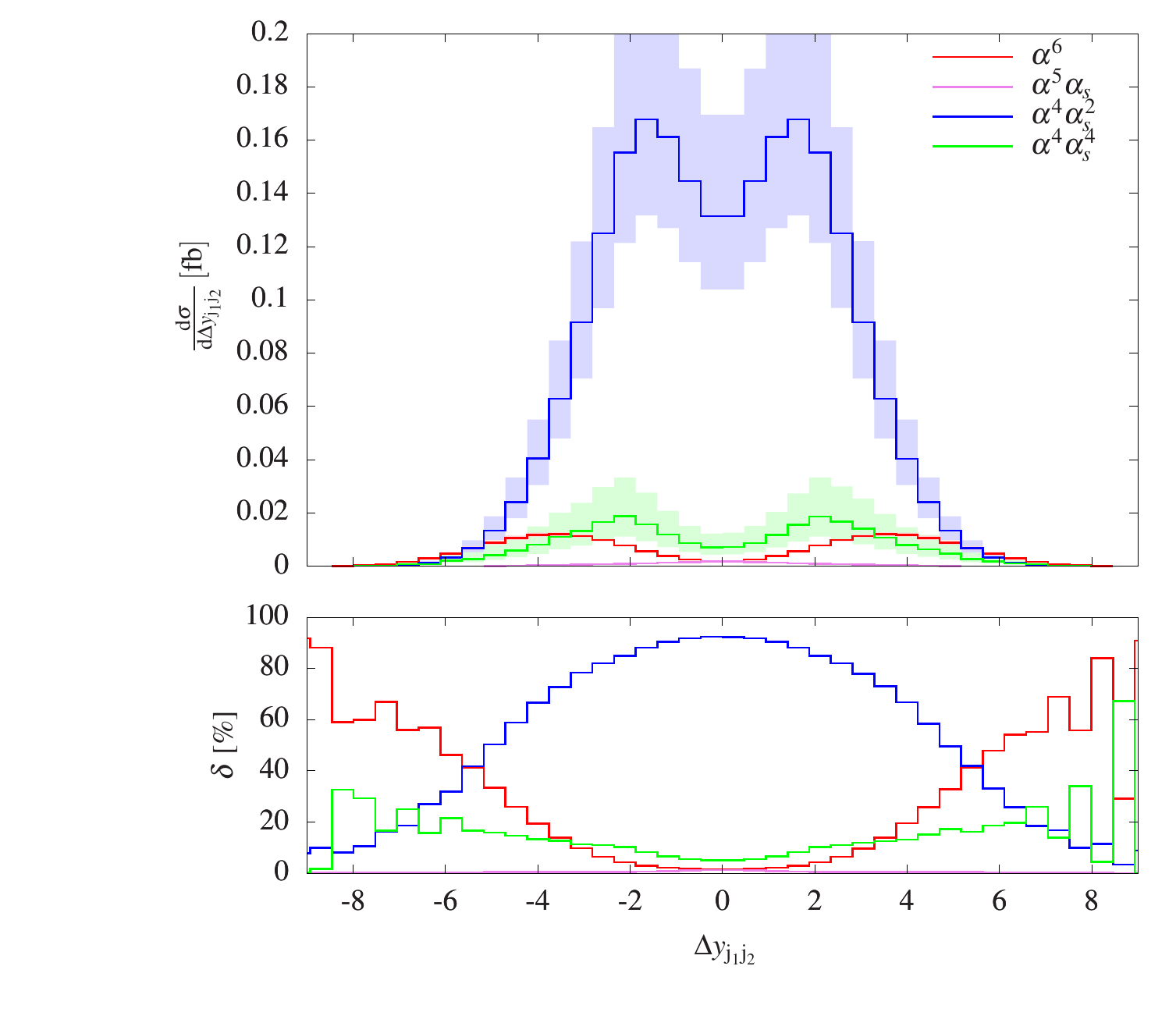}
\label{fig:lo_deltayjj} 
\end{subfigure}
\begin{subfigure}{0.49\textwidth}
\centering
\subcaption{}
\includegraphics[width=1.\linewidth]{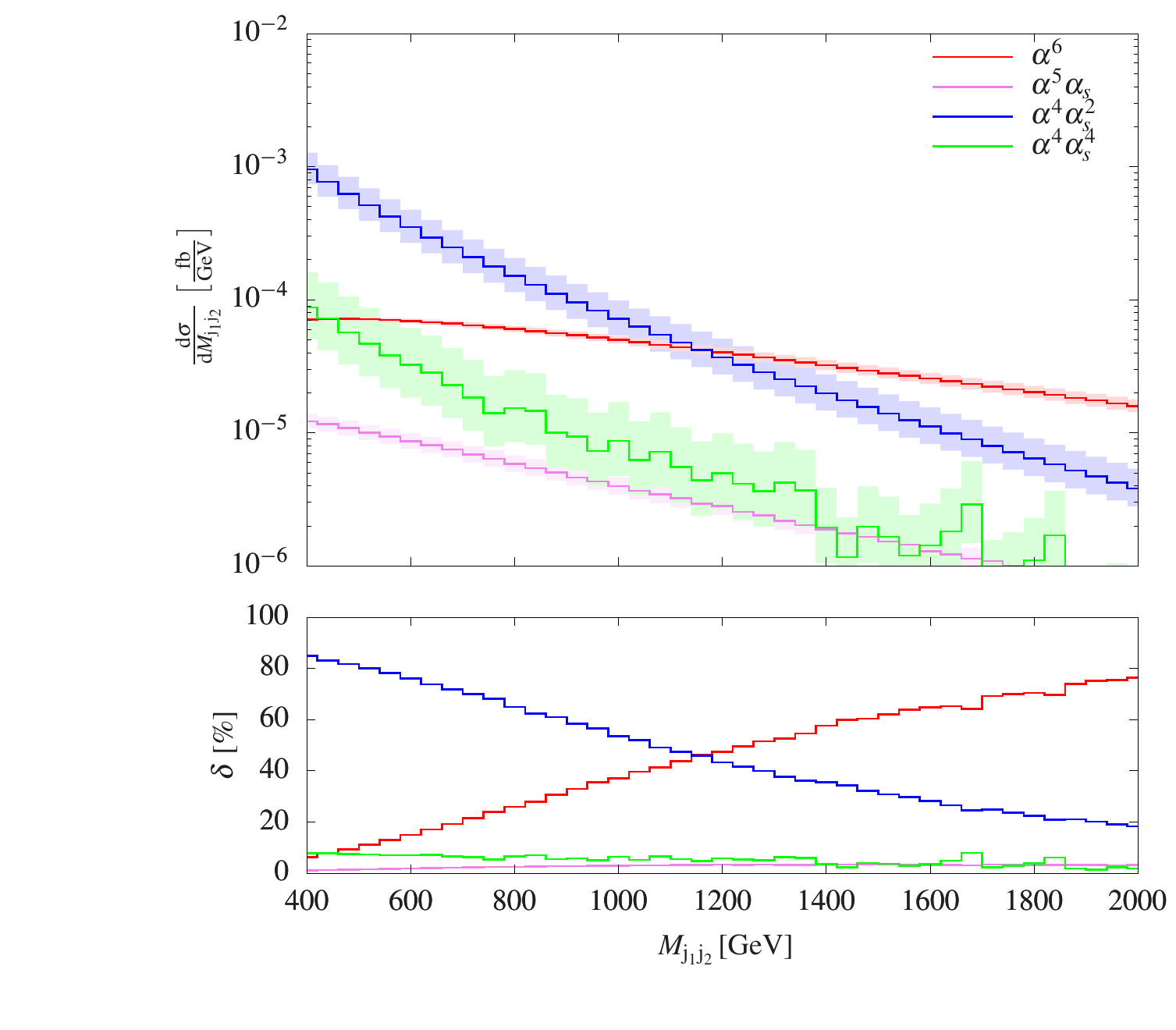}
\label{fig:lo_mjj}
\end{subfigure}%
\par\bigskip
\begin{subfigure}{0.49\textwidth}
\centering
\subcaption{}
\includegraphics[width=1.\linewidth]{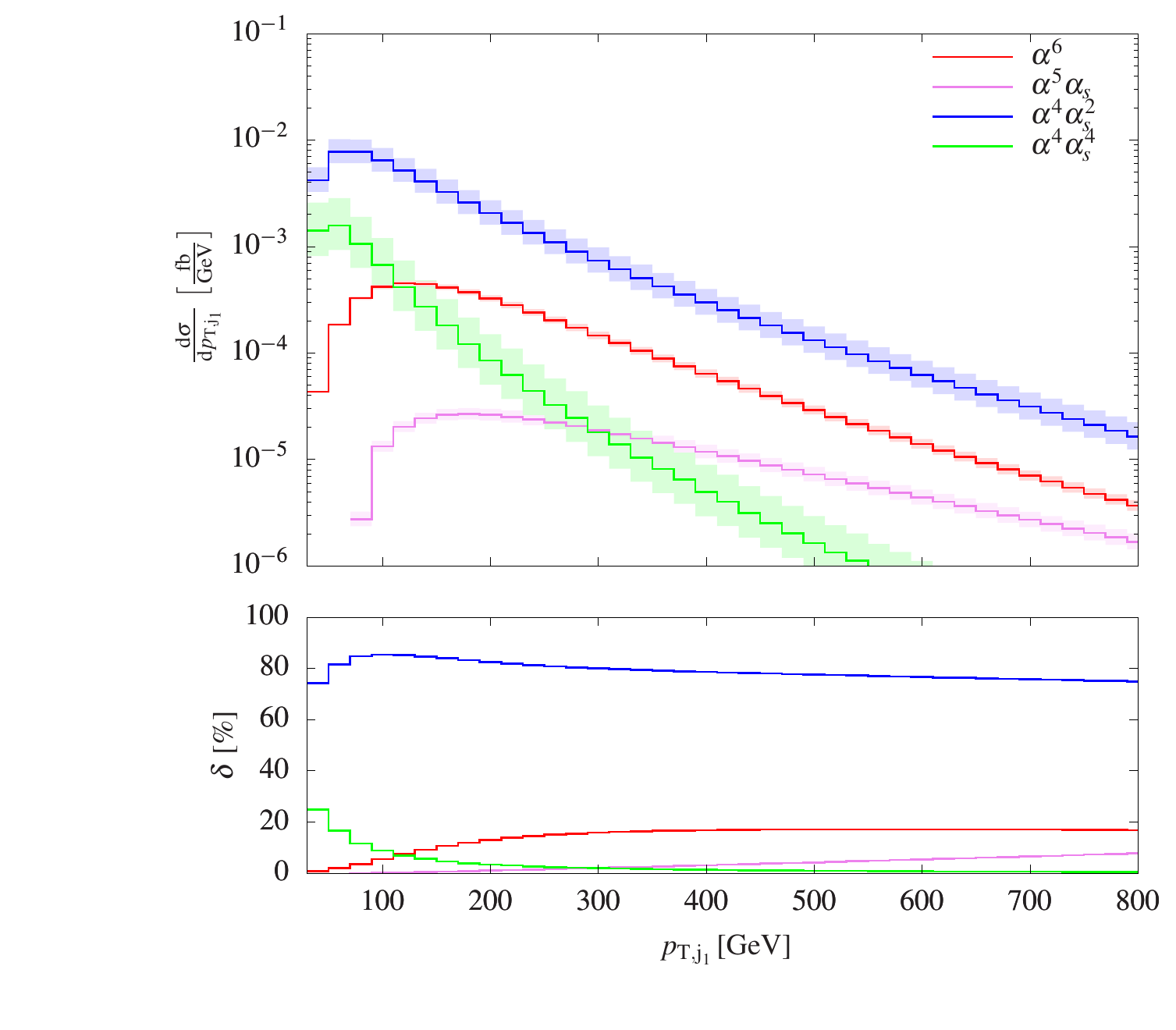}
\label{fig:lo_pTj1}
\end{subfigure}%
\begin{subfigure}{0.49\textwidth}
\centering
\subcaption{}
\includegraphics[width=1.\linewidth]{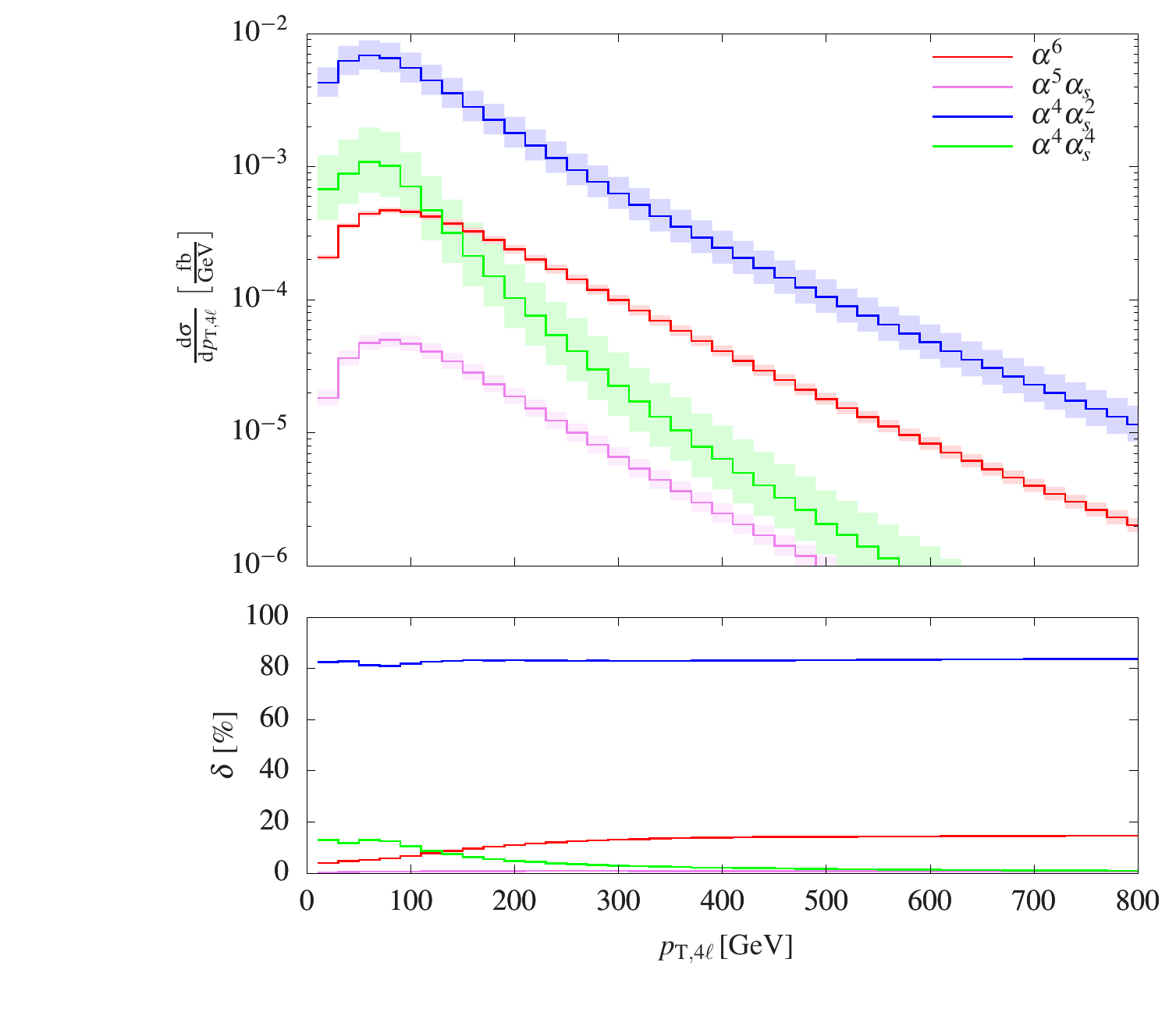}
\label{fig:lo_pTzz} 
\end{subfigure}
\vspace*{-3ex}
\caption{LO differential distributions at orders $\order{\alpha^6 }$ (EW contribution), 
$\order{\alphas \alpha^5 }$ (interference), and $\order{\alphas^2 \alpha^4 }$ (QCD contribution)
combined with the loop-induced contribution of order $\order{\alphas^4 \alpha^4 }$.
The upper panels show absolute predictions, while the lower ones show
each contribution relative to the sum of all of them. 
The observables read as follows:
rapidity difference between the two hardest jets (top left),
invariant mass of the two hardest jets (top right),
transverse momentum of the hardest jet (bottom left),
and
transverse momentum of the 4 leptons (bottom right).}
\label{fig:LO}
\end{figure}

First, we display distributions at LO in \reffi{fig:LO} including
theoretical predictions at orders $\order{\alpha^6 }$, $\order{\alphas
  \alpha^5 }$, and $\order{\alphas^2 \alpha^4 }$.  
In addition, we include predictions for the loop-induced contributions
at the order $\order{\alphas^4 \alpha^4 }$.  While the upper panels show the
absolute predictions, the lower ones display each contribution
relative to the sum of all of them.

The rapidity difference between the two hardest jets, shown in
Figure~\ref{fig:lo_deltayjj}, is a typical handle to enhance the EW
contribution.  As expected, the EW contribution becomes dominant only
in a rather extreme part of the phase space, typically for $|\Delta
y_{\Pj_1\Pj_2}| > 5$.  The central region is largely dominated by the
QCD contributions which amount to more than $90\%$ of the total
process.  The interference contribution is below $1\%$ and hardly
visible in the figure. The loop-induced process is particularly
interesting.  It is minimal for low rapidity
difference but relatively increasing towards large rapidity difference
like the EW contribution.

A further variable that is used to enhance the EW contribution is the di-jet
invariant mass, whose distribution is shown in \reffi{fig:lo_mjj}.
The EW contribution exceeds the QCD contribution only for
$M_{\Pj_1\Pj_2}>1200\GeV$. The fraction of the loop-induced
contribution decreases with increasing invariant mass, while the
interference increases but stays below $5\%$ for $M_{\Pj_1\Pj_2}<2000\GeV$.

In \reffi{fig:lo_pTj1}, the distribution in the transverse momentum of
the hardest jet is shown.  The relative EW contribution is slowly
increasing with $p_{\rT,\Pj_1}$ to reach almost $20\%$ at $400\GeV$.
The interference contribution becomes non negligible in this part of
phase space and amounts to about $10\%$ for $p_{\rT,\Pj_1}=800\GeV$.
Finally, the loop-induced contribution drops quickly and
is below $3\%$
above $200\GeV$.

The distribution in the transverse momentum of the four leptons,
shown in \reffi{fig:lo_pTzz}, behaves qualitatively similar to the
distribution in the transverse momentum of the hardest jet.  This is
expected as these two observables are correlated.  It is worth
noticing that the interference contribution does not increase towards
high transverse momentum as in the previous case and is thus almost
imperceptible over the whole range.  Also, the loop-induced
contribution is dropping less quickly than in the previous case.  In
particular, in the first bin of the distribution in the transverse
momentum of the hardest jet, the loop-induced process represents about
$25\%$ of the total predictions, while here it is about $15\%$.

\subsubsection*{NLO distributions}

We turn to distributions including NLO corrections. In the
following figures, the upper panels show the absolute predictions for
the LO EW component of $\order{\alpha^6 }$ complemented by predictions
including the orders $\order{\alpha^7 }$ (NLO EW) or $\order{\alphas
  \alpha^6 }$ (NLO QCD).  In addition, the best prediction is denoted
NLO EW+QCD and includes both types of corrections summed.  In the
lower panel, the three NLO predictions are normalised to the LO EW
predictions of $\order{\alpha^6 }$.

\begin{figure}
\setlength{\parskip}{-4ex}
\begin{subfigure}{0.49\textwidth}
\centering
\subcaption{}
\includegraphics[width=1.\linewidth]{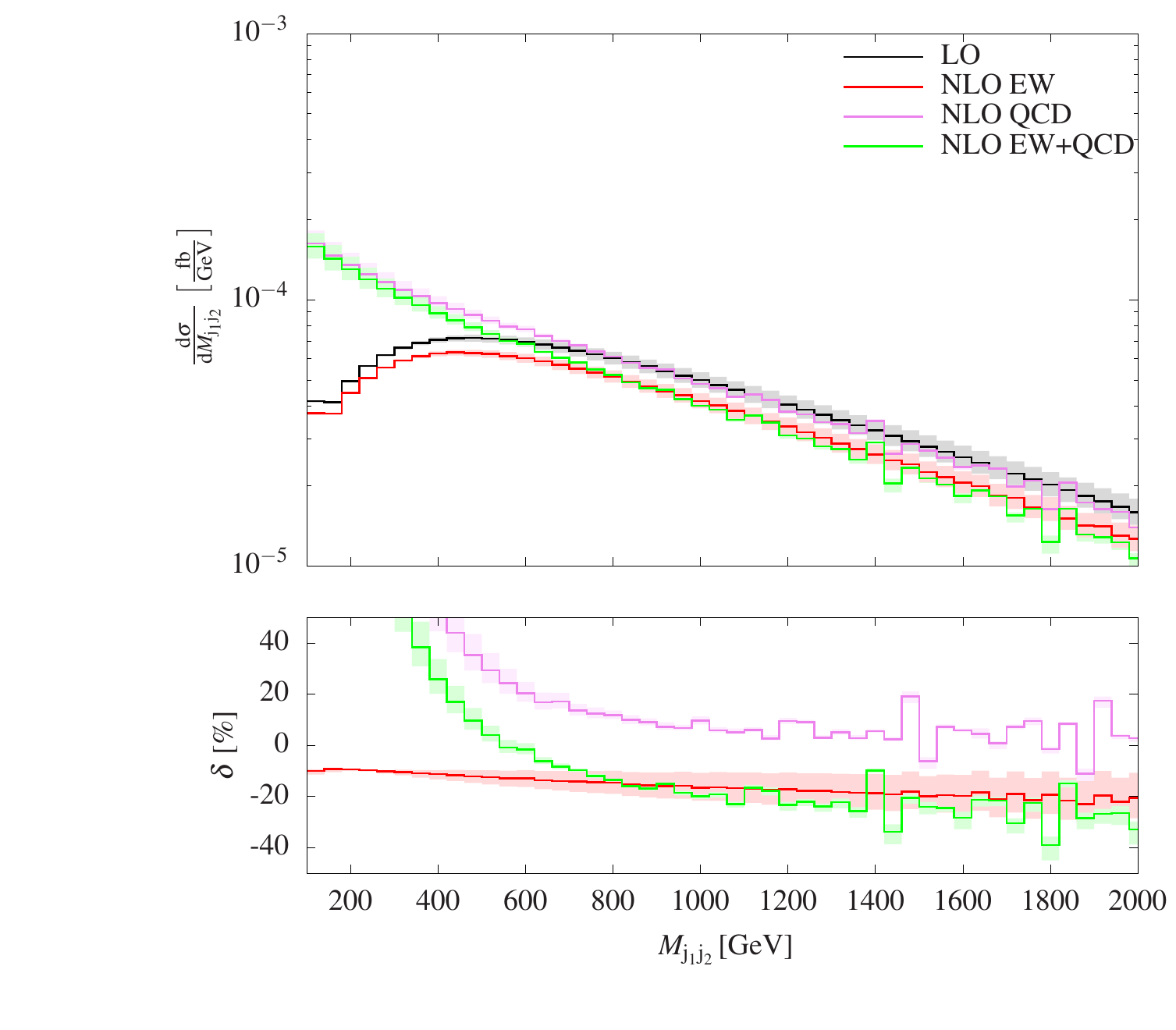}
\label{fig:mjj} 
\end{subfigure}
\begin{subfigure}{0.49\textwidth}
\centering
\subcaption{}
\includegraphics[width=1.\linewidth]{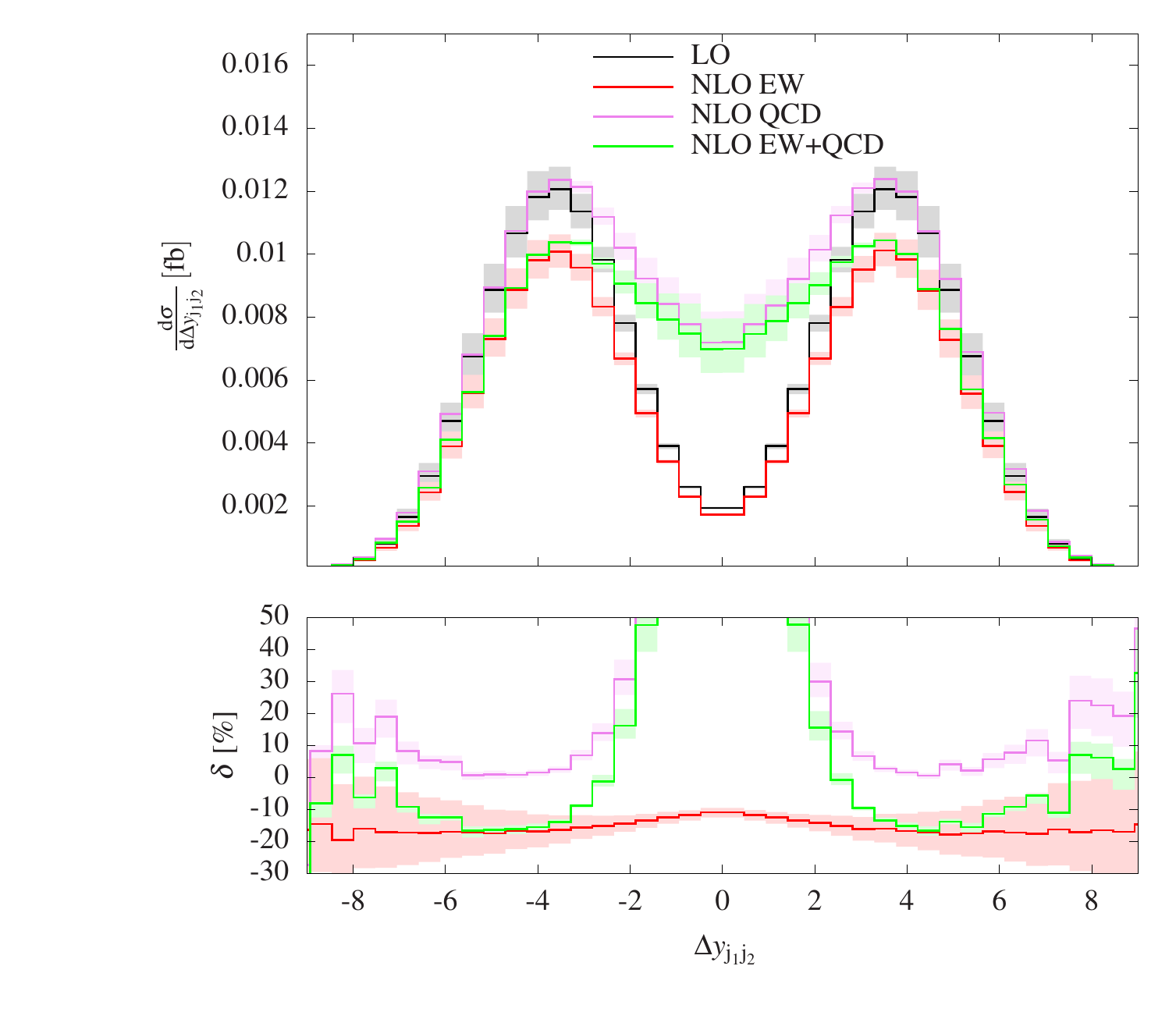}
\label{fig:dyjj}
\end{subfigure}%
\par\bigskip
\begin{subfigure}{0.49\textwidth}
\centering
\subcaption{}
\includegraphics[width=1.\linewidth]{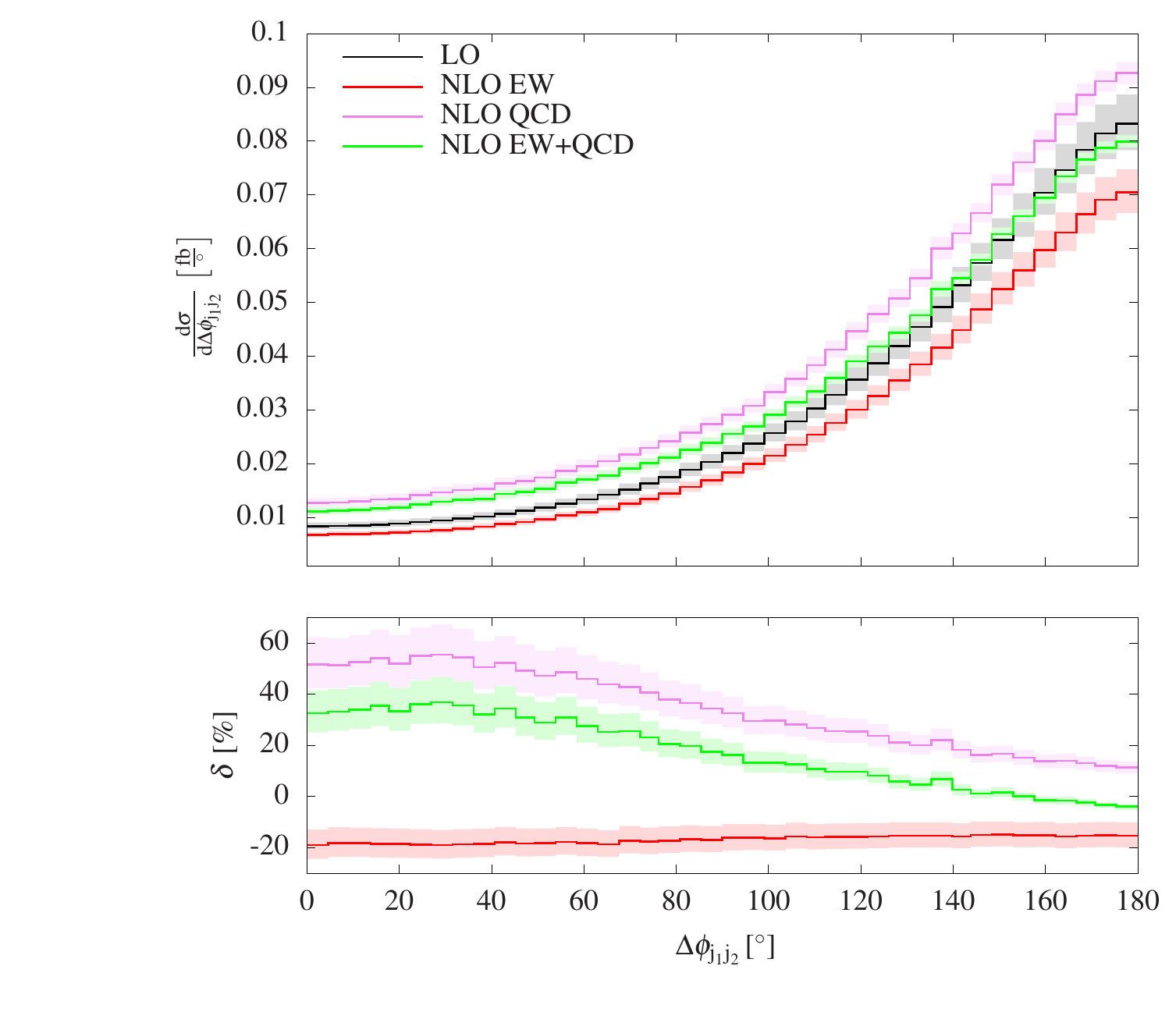}
\label{fig:phijj} 
\end{subfigure}
\begin{subfigure}{0.49\textwidth}
\centering
\subcaption{}
\includegraphics[width=1.\linewidth]{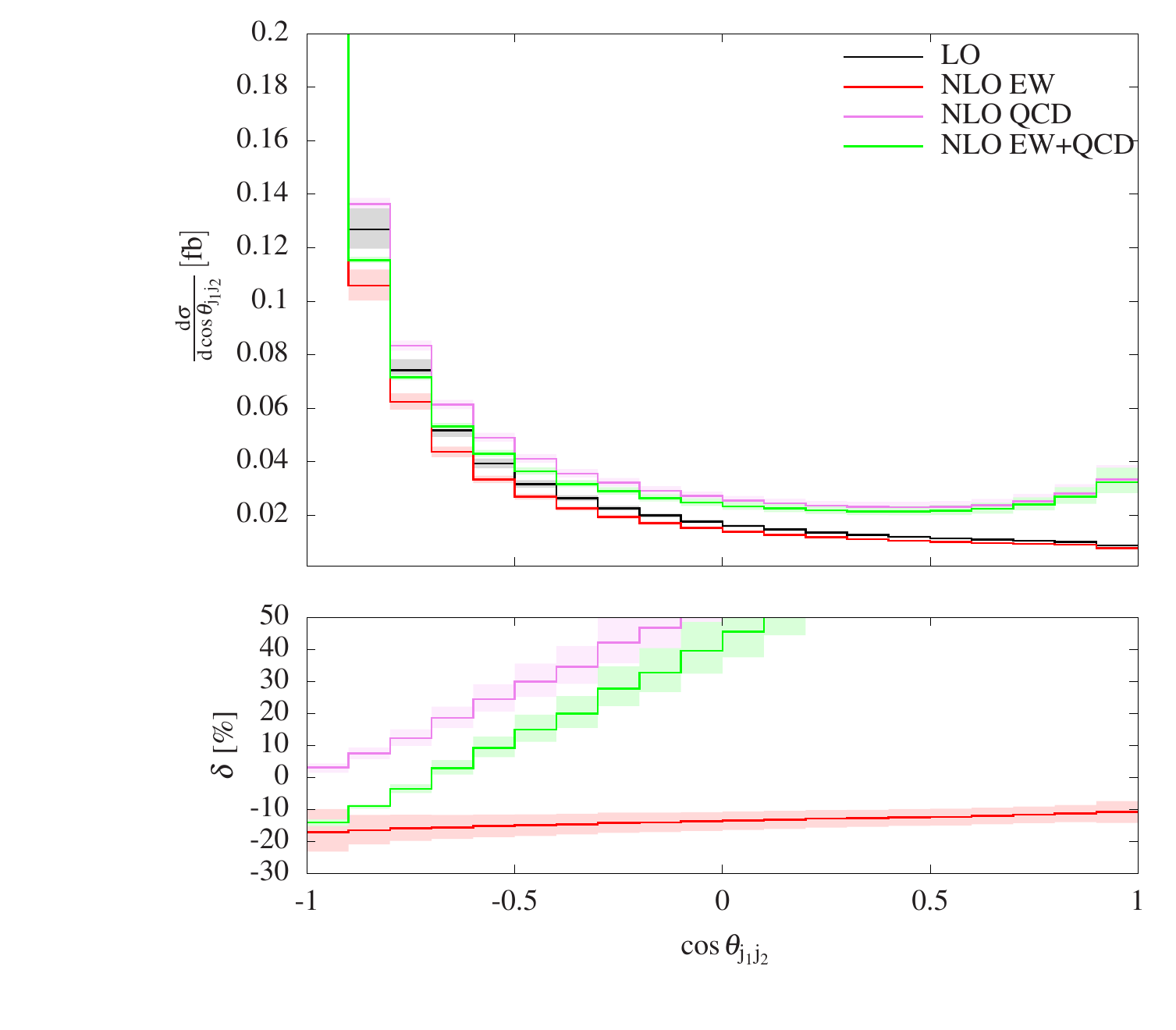}
\label{fig:cosjj}
\end{subfigure}%
\vspace*{-3ex}
\caption{LO and NLO differential distributions at orders $\order{\alpha^6}$ (LO), 
$\order{\alpha^7}$ (NLO EW), $\order{\alphas\alpha^6}$ (NLO QCD), and NLO EW+QCD.
The upper panels show absolute predictions while the lower ones show
each contribution relative to the LO predictions. 
The observables read as follows:
invariant mass of the two hardest jets (top left),
rapidity separation of the two hardest jets (top right),
azimuthal angle between the two hardest jets (bottom left), and
cosine of the angle between the two hardest jets (bottom right).}
\label{fig:NLOjj}
\end{figure}

The observables shown in \reffi{fig:NLOjj} are related to the
two hardest jets.  We start with the two observables that are typically
used to enhance EW contributions over its irreducible QCD background:
the invariant mass (\reffi{fig:mjj}) and the rapidity difference
(\reffi{fig:dyjj}) of the two leading jets.  
The most interesting feature is that the bulk of the positive QCD
corrections is located at low di-jet invariant mass.  The
$\order{\alphas \alpha^6 }$ corrections are well above $100\%$ in the
first bins but become of the order of a few percent above $600\GeV$.
This effect has been already observed and partially discussed in
\citere{Ballestrero:2018anz} for like-sign W-boson scattering and is
due the appearance of W or Z bosons becoming resonant and decaying to
a pair of jets when including real gluon radiation.  As shown in
\reffis{fig:born_qq_triW} and \ref{fig:born_qq_triZga},
$\Pp\Pp\to\Pe^{+}\Pe^{-}\mu^{+}\mu^{-}\Pj\Pj$ also includes
contributions from triple vector-boson production ($\PW\PZ\PZ$ and
$\PZ\PZ\PZ$) at order $\order{\alpha^6 }$.  At LO, the massive vector
boson decaying hadronically cannot become resonant due to the cut on
the invariant mass of the two jets at $100\GeV$ [see
Eq.~\eqref{eq:vbscuts}].  When including real radiation at NLO, the
cut \eqref{eq:vbscuts} does not necessarily apply to the two quarks
coming from the vector-boson decay.  An extra gluon jet can make up
the $M_{\Pj_1\Pj_2} > 100\GeV$ and allows the two quarks to originate
from a resonant W or Z~boson.  Thus, the relatively large QCD
corrections found here are a combined effect of the real radiation,
the event selection, and the inclusion of tri-boson contributions.  We
would like to emphasise that the phase-space region of our default
setup is rather inclusive and should be avoided if one does not
include tri-boson contributions in the theory predictions.  Otherwise,
the fiducial corrections at NLO QCD will most likely be off by about
$20\%$, and this offset is by no means accounted for by
scale-variation uncertainties.  This consideration is particularly
important if Monte Carlo programs using the VBS approximation
\cite{Ballestrero:2018anz} are used to extrapolate measurements from
the inclusive region to more VBS-enriched regions.  An alternative is
to subtract the on-shell tri-boson contributions from the results for
the full process.  Such a strategy is often used in experimental
analysis to extract VBS contributions.  While it allows one to claim a
\emph{VBS measurement}, care has to be taken not to violate gauge
invariance.  Furthermore, it has the disadvantage to make the
measurement even more theory dependent.  In our opinion, the most
straightforward and physical measurement would include both the QCD
and the EW processes and in the latter all possible contributions
including tri-boson production, \ie all contributions to a given
physical final state.
The EW corrections, on the other hand, do not display an unexpected
behaviour but confirm the results known from other VBS signatures
\cite{Biedermann:2016yds,Biedermann:2017bss,Denner:2019tmn}.  They
become negatively large for large invariant masses owing to enhanced
EW logarithms to reach $-20\%$ at $2\TeV$.

Turning to the distribution in the rapidity difference shown in
\reffi{fig:dyjj}, the QCD corrections reach almost
$300\%$ in the central rapidity region. The rapidity
  separation of the two hardest jets is strongly correlated to their
  invariant mass (see, for instance, Figure 3 of
  \citere{Ballestrero:2018anz}). Thus, the arguments given for the
  distribution in $M_{\Pj_1\Pj_2}$ can be transfered to the
  distribution in $\Delta y_{\Pj_1\Pj_2}$. Events with small
  $\Delta y_{\Pj_1\Pj_2}$ are depleted at LO owing to the cut
  \eqref{eq:vbscuts}, while this is not the case at NLO QCD where
  extra gluons can provide a leading jet.  The distribution also shows
  that a cut on the rapidity difference would be very effective in
  removing the sizeable QCD corrections linked to triple-vector-boson
  production in a similar way as a stronger cut on $M_{\Pj_1\Pj_2}$.
Thus, the large QCD corrections could be reduced by either a cut on
$\Delta y_{\Pj_1\Pj_2}$ or a stronger one on $M_{\Pj_1\Pj_2}$, which are
usually imposed in VBS studies.
  The EW corrections are moderate and vary between $-10\%$ for zero
  rapidity difference
and $-20\%$ for large rapidity differences.

Figures~\ref{fig:phijj} and \ref{fig:cosjj} show distributions in the
azimuthal-angle difference and the cosine of the angle between the two
leading jets, respectively, which provide information on the
correlation between the two jets.  The EW corrections are rather
stable throughout the kinematic range and vary by less than $7\%$.
For the azimuthal-angle difference, the QCD corrections are maximal
near $30^\circ$ where they are about $55\%$.  
When the two jets have
maximal azimuthal-angle difference, the LO contribution is maximal and
receives QCD corrections at the level of $10\%$. 
 The distribution in
$\cos\theta_{\Pj_1\Pj_2}$ peaks at $-1$, \ie when the two jets are
back-to-back.  The QCD corrections are minimal there but exceed
$200\%$ when the two jets are close to each other.

\begin{figure}
\setlength{\parskip}{-4ex}
\begin{subfigure}{0.49\textwidth}
\centering
\subcaption{}
\includegraphics[width=1.\linewidth]{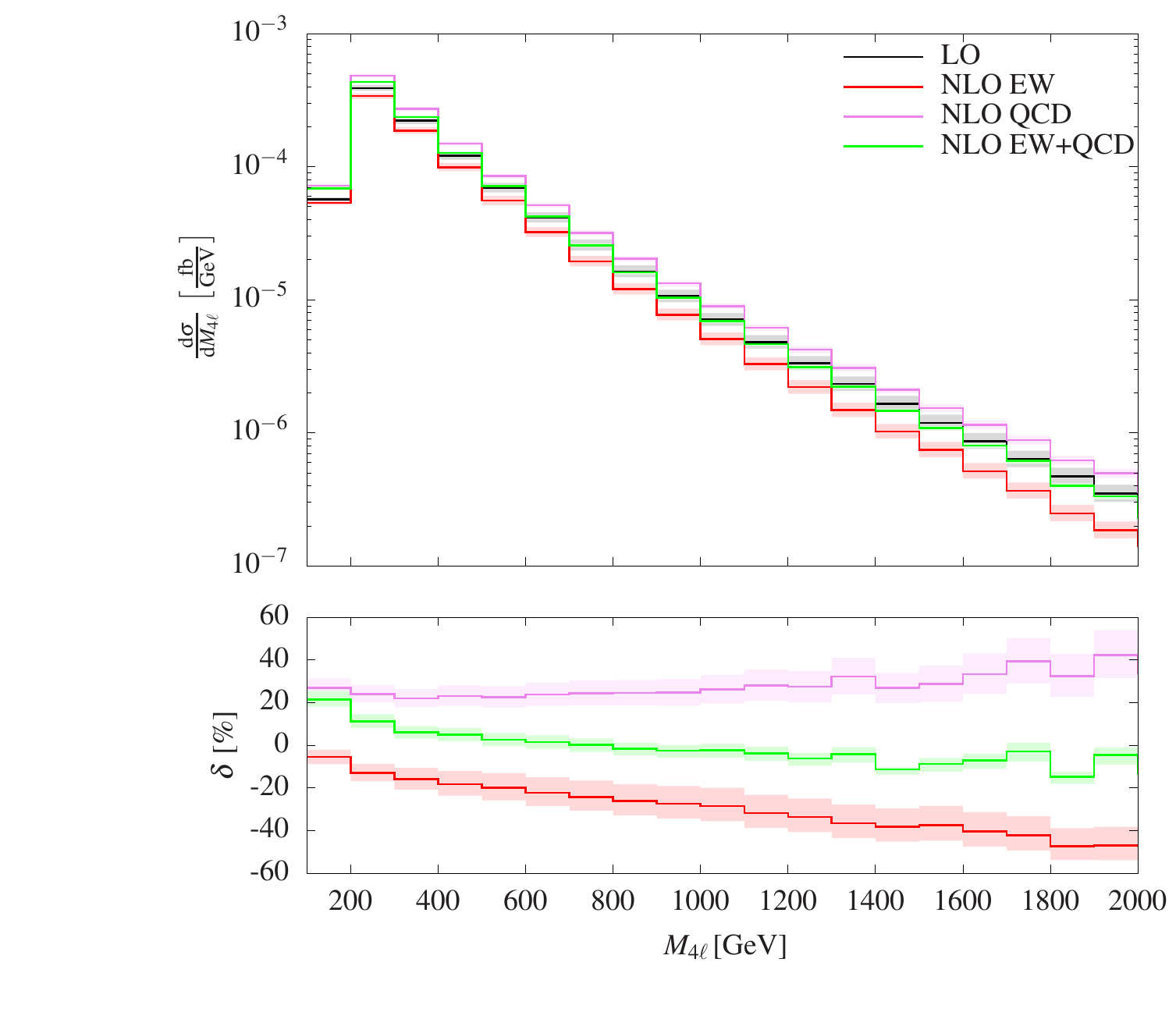}
\label{fig:m4l} 
\end{subfigure}
\begin{subfigure}{0.49\textwidth}
\centering
\subcaption{}
\includegraphics[width=1.\linewidth]{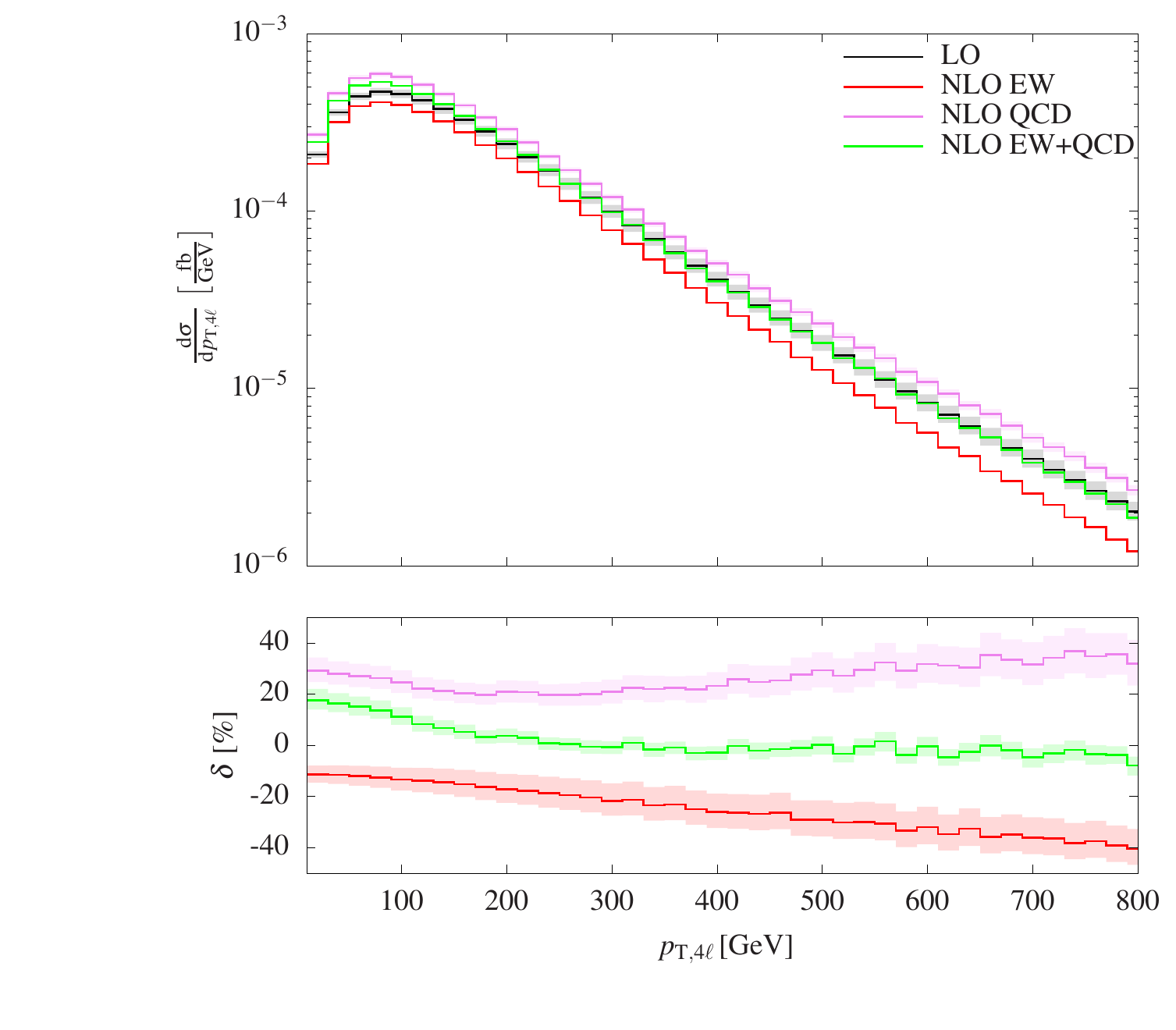}
\label{fig:pTzz} 
\end{subfigure}
\par\bigskip
\begin{subfigure}{0.49\textwidth}
\centering
\subcaption{}
\includegraphics[width=1.\linewidth]{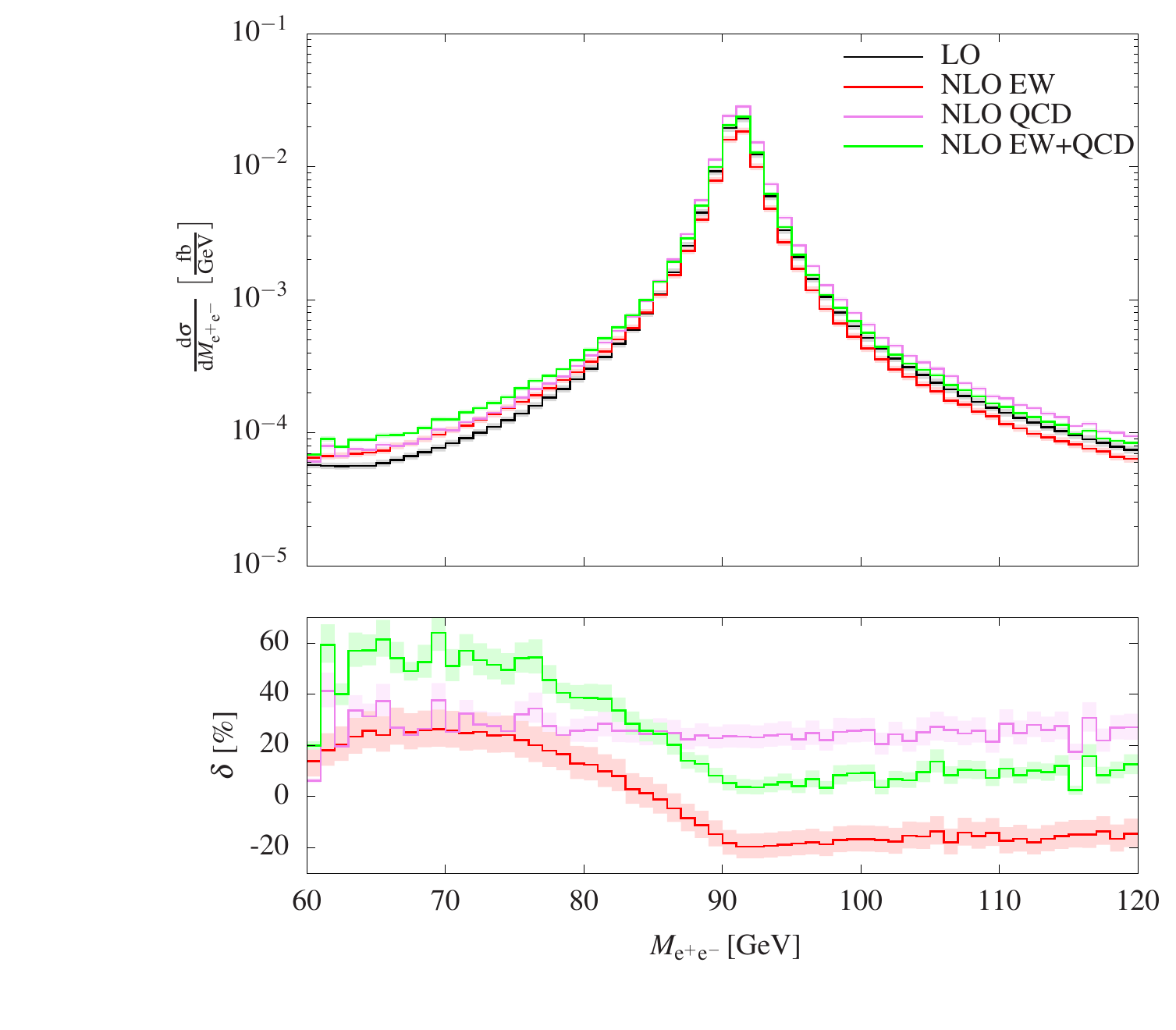}
\label{fig:mee}
\end{subfigure}%
\begin{subfigure}{0.49\textwidth}
\centering
\subcaption{}
\includegraphics[width=1.\linewidth]{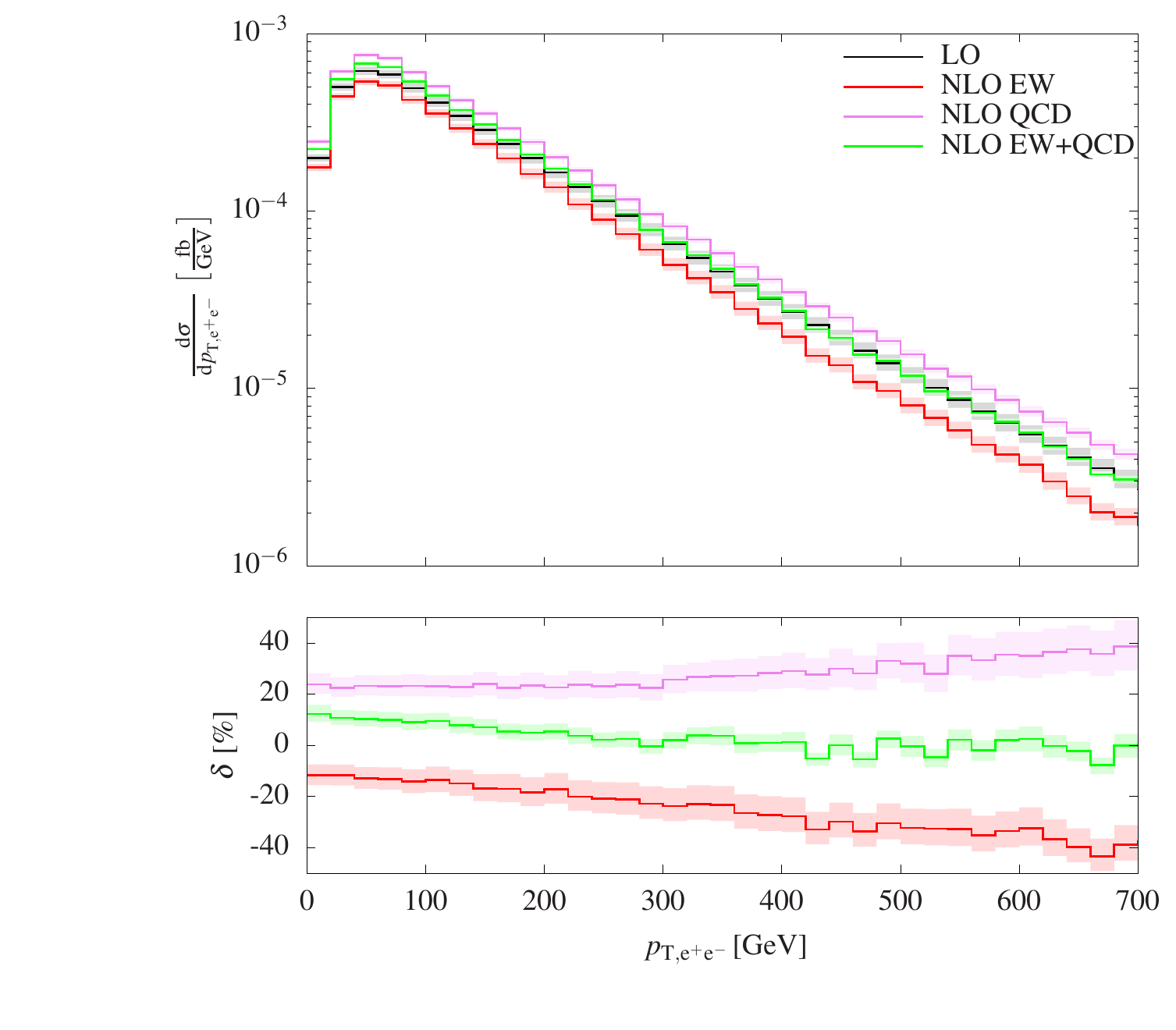}
\label{fig:pTee}
\end{subfigure}%
\vspace*{-3ex}
\caption{LO and NLO differential distributions at orders $\order{\alpha^6 }$ (LO), 
$\order{\alpha^7 }$ (NLO EW), $\order{\alphas\alpha^6 }$ (NLO QCD), and NLO EW+QCD.
The upper panels show absolute predictions while the lower ones show
each contribution relative to the LO predictions. 
The observables read as follows:
invariant mass of 4-lepton system (top left),
transverse momentum of the 4 leptons (top right),
invariant mass of the electron--positron system (bottom left), 
and
transverse momentum of the electron--positron system (bottom right).}
\label{fig:NLOZ}
\end{figure}

In \reffi{fig:NLOZ} we display distributions related to the 4-lepton
system, \ie the \PZ-boson pair, and the electron--positron pair, \ie
one of the \PZ~bosons.
The distribution in the invariant mass of the 4-lepton system receives
the typical EW Sudakov corrections growing to $-40\%$ at
$M_{4\Pl}=2\TeV$. The QCD corrections increase from $20\%$ to $40\%$
in the considered range. As a consequence, the overall NLO corrections
experience a cancellation for larger invariant masses and decrease
from $20\%$ to almost zero towards large $M_{4\Pl}$.
The distribution in the transverse momentum of the 4 leptons
(\reffi{fig:pTzz}) is particularly interesting
as it is the transverse momentum of the vector-boson scattering
subprocess.  The corrections behave similarly as for the distribution
in the invariant mass of the 4-lepton system and show cancellations
for large transverse momenta.  The EW corrections steadily increase to
$-40\%$ at $\ptsub{4\Pl}=800\GeV$. The relative QCD ones, on the
other hand, reach a minimum near $200\GeV$ and slowly increase up to
$40\%$ at $800\GeV$.  
The distribution in the invariant mass of the electron--positron
system, presented in \reffi{fig:mee}, shows a typical Z-boson
resonance.  While the QCD corrections hardly modify the shape of this
distribution, the EW ones exhibit a radiative tail below the resonance
reaching more than $+20\%$.  This tail is due to real photon radiation
that takes away part of the energy of the electron--positron system
and thus shifts events from the peak towards lower invariant masses.
Such an effect is well known and has been observed already for
Drell--Yan, di-boson, or top--anti-top production processes.
The distribution in the transverse momentum of the electron--positron
system, \ie the transverse momentum of one of the Z~bosons, is
presented in \reffi{fig:pTee}. The EW corrections increase from
$-10\%$ to $-40\%$ and the QCD corrections from $25\%$ to
$40\%$, leading to an overall NLO correction decreasing from $15\%$ to
almost zero.

\begin{figure}
\setlength{\parskip}{-4ex}
\begin{subfigure}{0.49\textwidth}
\centering
\subcaption{}
\includegraphics[width=1.\linewidth]{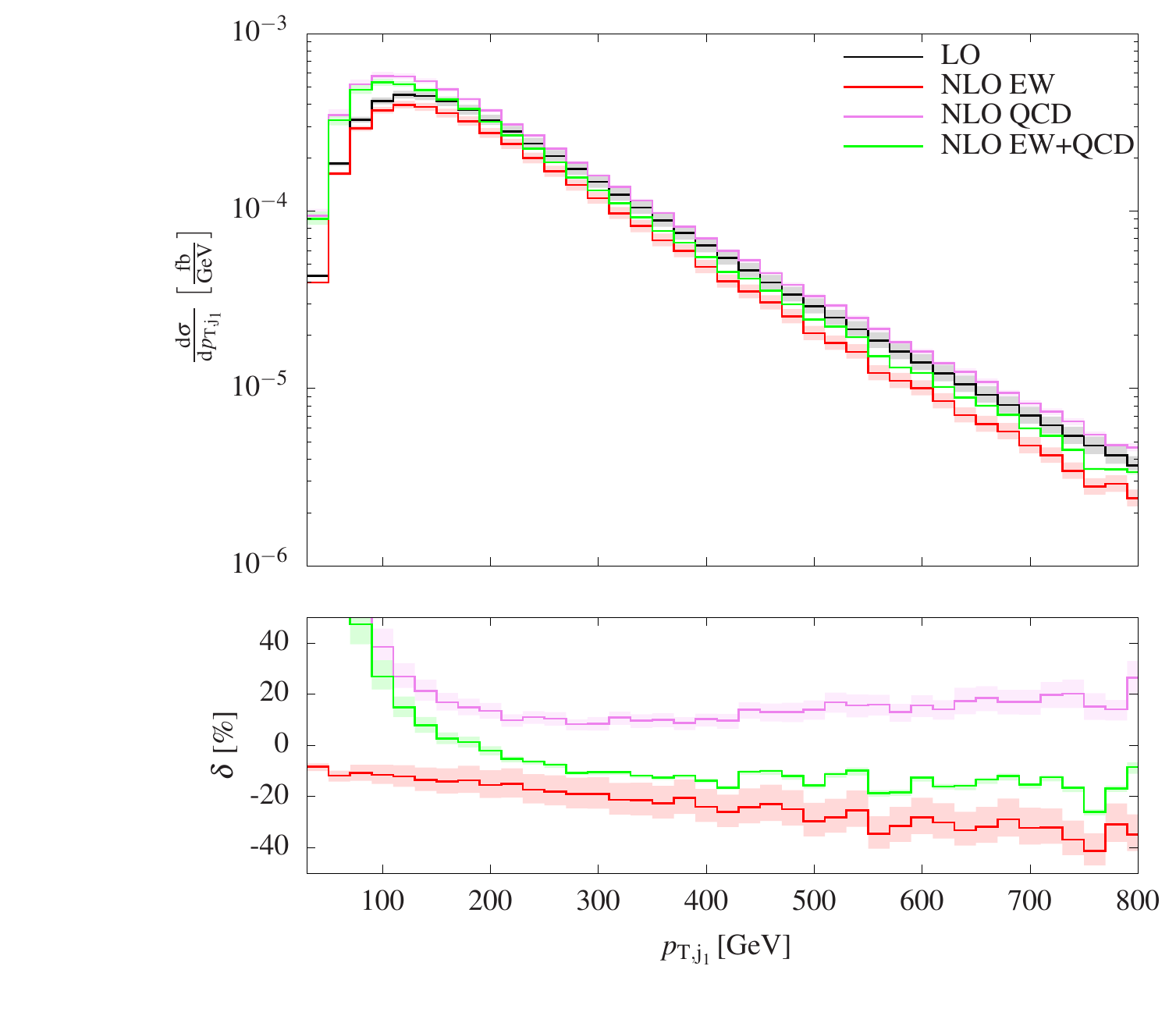}
\label{fig:pTj1} 
\end{subfigure}
\begin{subfigure}{0.49\textwidth}
\centering
\subcaption{}
\includegraphics[width=1.\linewidth]{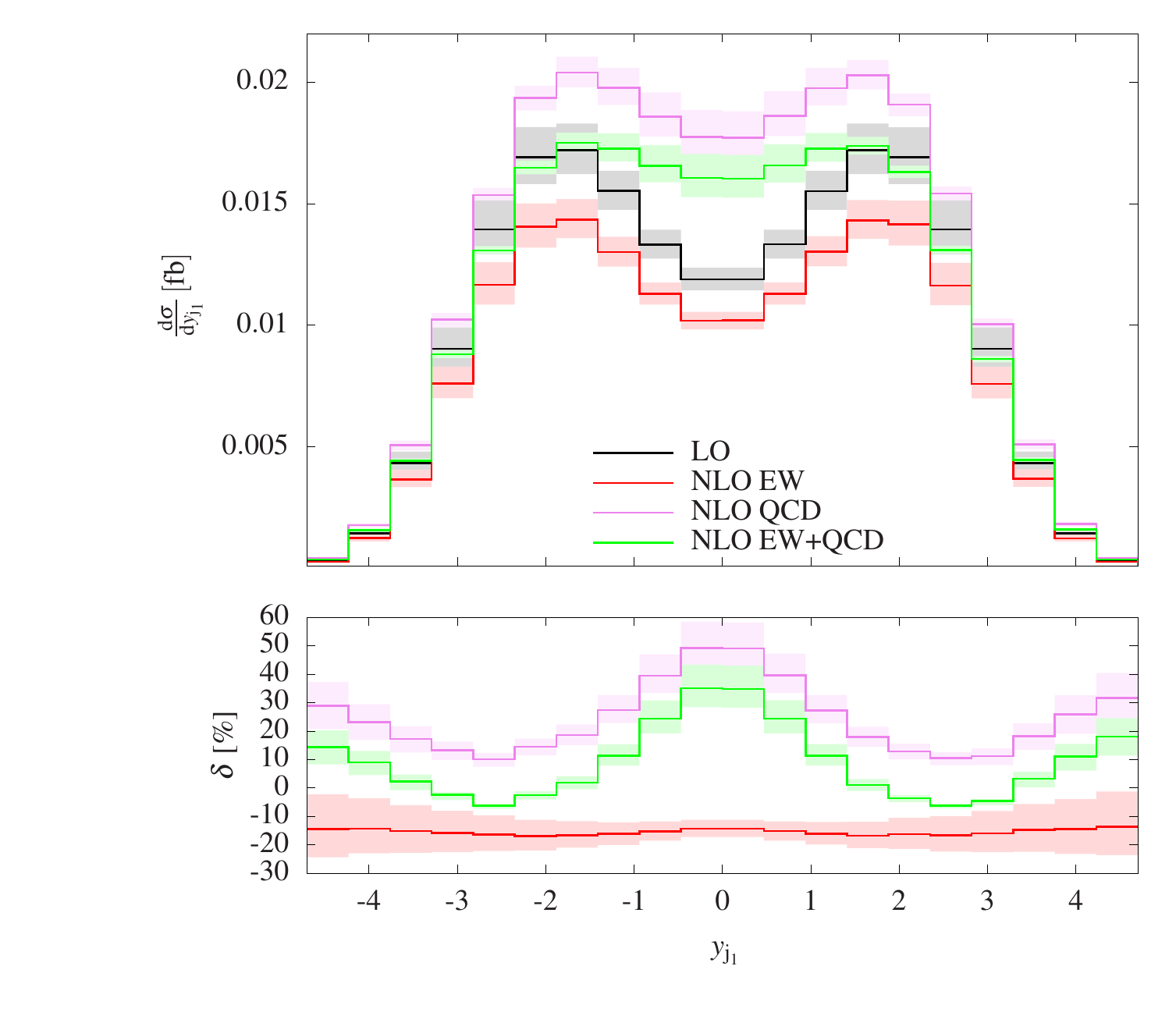}
\label{fig:yj1}
\end{subfigure}%
\par\bigskip
\begin{subfigure}{0.49\textwidth}
\centering
\subcaption{}
\includegraphics[width=1.\linewidth]{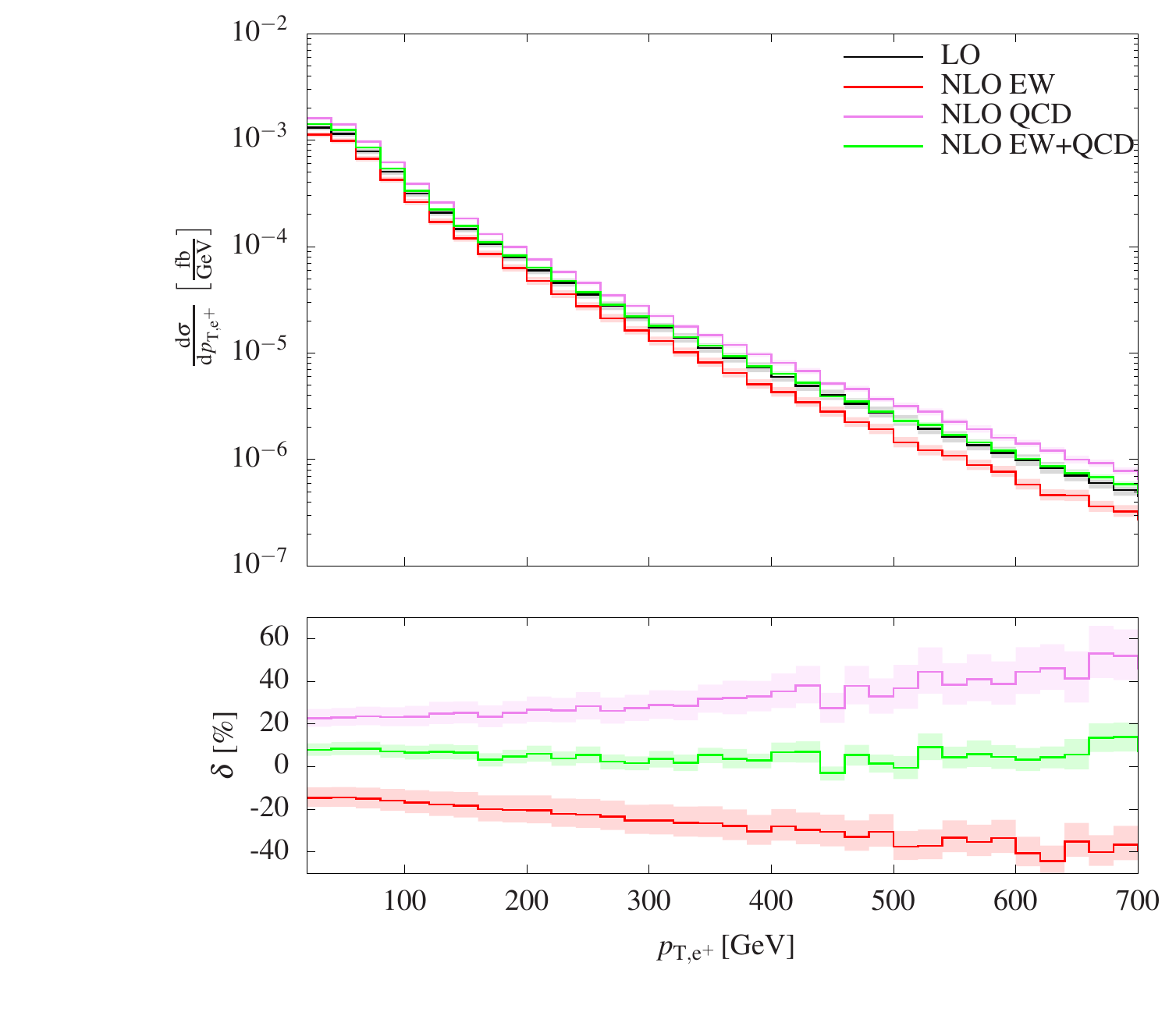}
\label{fig:pTep} 
\end{subfigure}
\begin{subfigure}{0.49\textwidth}
\centering
\subcaption{}
\includegraphics[width=1.\linewidth]{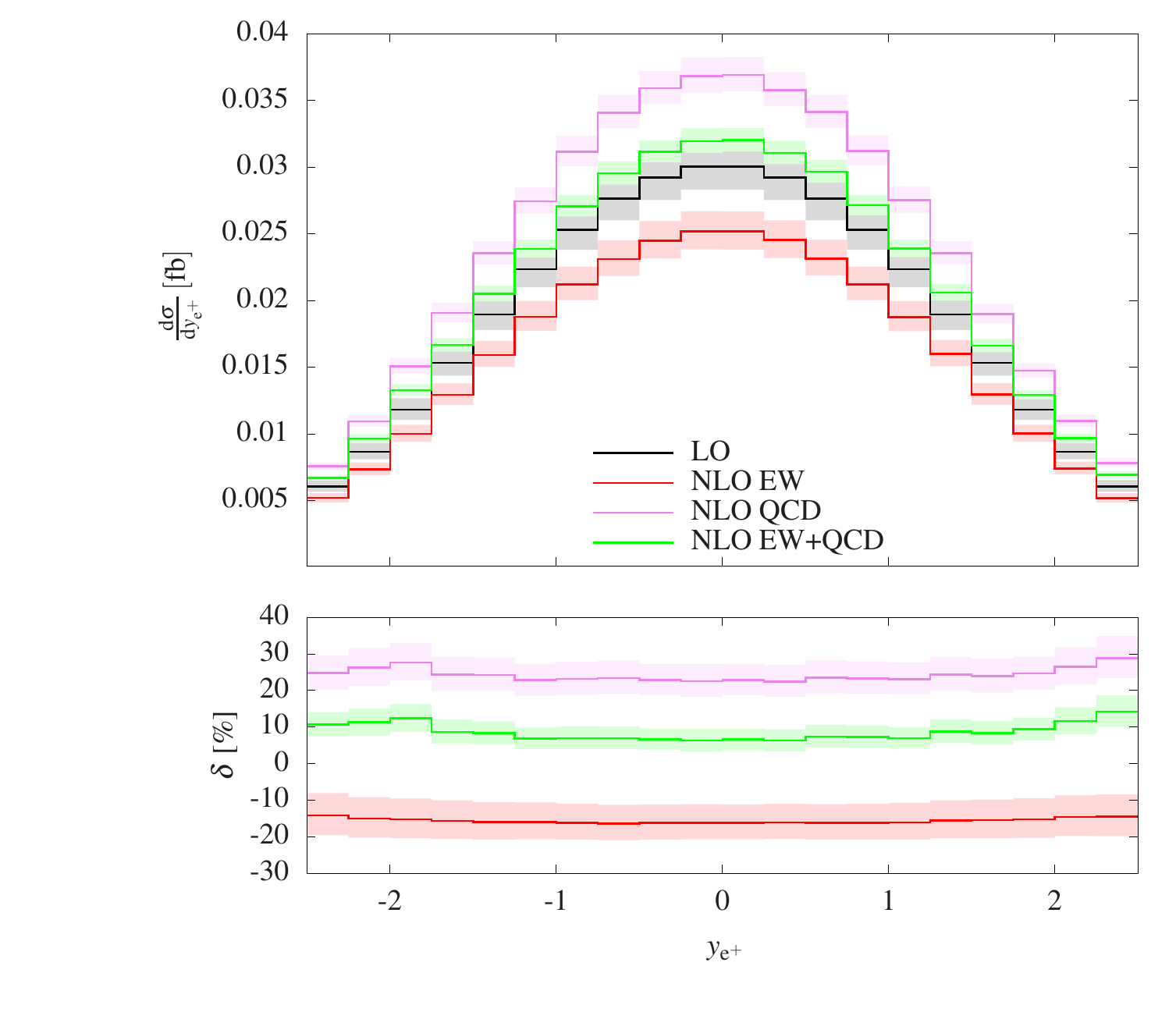}
\label{fig:yep}
\end{subfigure}%
\vspace*{-3ex}
\caption{LO and NLO differential distributions at orders $\order{\alpha^6 }$ (LO), 
$\order{\alpha^7 }$ (NLO EW), $\order{\alphas\alpha^6 }$ (NLO QCD), and NLO EW+QCD.
The upper panels show absolute predictions while the lower ones show
each contribution relative to the LO predictions. 
The observables read as follows:
transverse momentum of the hardest jet (top left),
rapidity of the hardest jet (top right),
transverse momentum of the positron (bottom left),
and
rapidity of the positron (bottom right).
}
\label{fig:NLOje}
\end{figure}

Next we study distributions in transverse momenta and rapidities of
the leading jet and the positron in \reffi{fig:NLOje}.
As observed in previous computations of VBS processes
\cite{Biedermann:2017bss,Denner:2019tmn}, the distribution in the
transverse momentum of the leading jet (\reffi{fig:pTj1}) is
suppressed for small transverse momenta and receives large QCD
corrections in this region.  This can be attributed to the presence of
an extra jet and reshuffling of energy between the jets.  In the rest
of the spectrum, the QCD corrections are at the level of $+20\%$ as
for the fiducial cross section.  The EW corrections show the typical
high-energy behaviour of other transverse-momentum distributions
growing negatively large with increasing transverse momentum and
reaching about $-40\%$ at $800\GeV$.  The distribution in the rapidity
of the hardest jet (\reffi{fig:yj1}) displays a similar behaviour as the
rapidity difference of the jet pair.  While the relative EW
corrections are rather flat, the QCD ones show large variations being
maximal in the central region and in the peripheral region.  At
intermediate rapidity, 
the corrections are the lowest with a bit more than $+10\%$.
The distribution in the transverse momentum of the positron
(\reffi{fig:pTep}) follows closely the distribution in the transverse
momentum of the electron--positron pair (\reffi{fig:pTee}), up to the
region of very small transverse momentum. Both the EW and QCD
corrections display the same behaviour for both distributions.
The distribution in the rapidity of the positron peaks in the central
region. Both QCD and EW corrections are flat.

\begin{figure}
\setlength{\parskip}{-4ex}
\begin{subfigure}{0.49\textwidth}
\centering
\subcaption{}
\includegraphics[width=1.\linewidth]{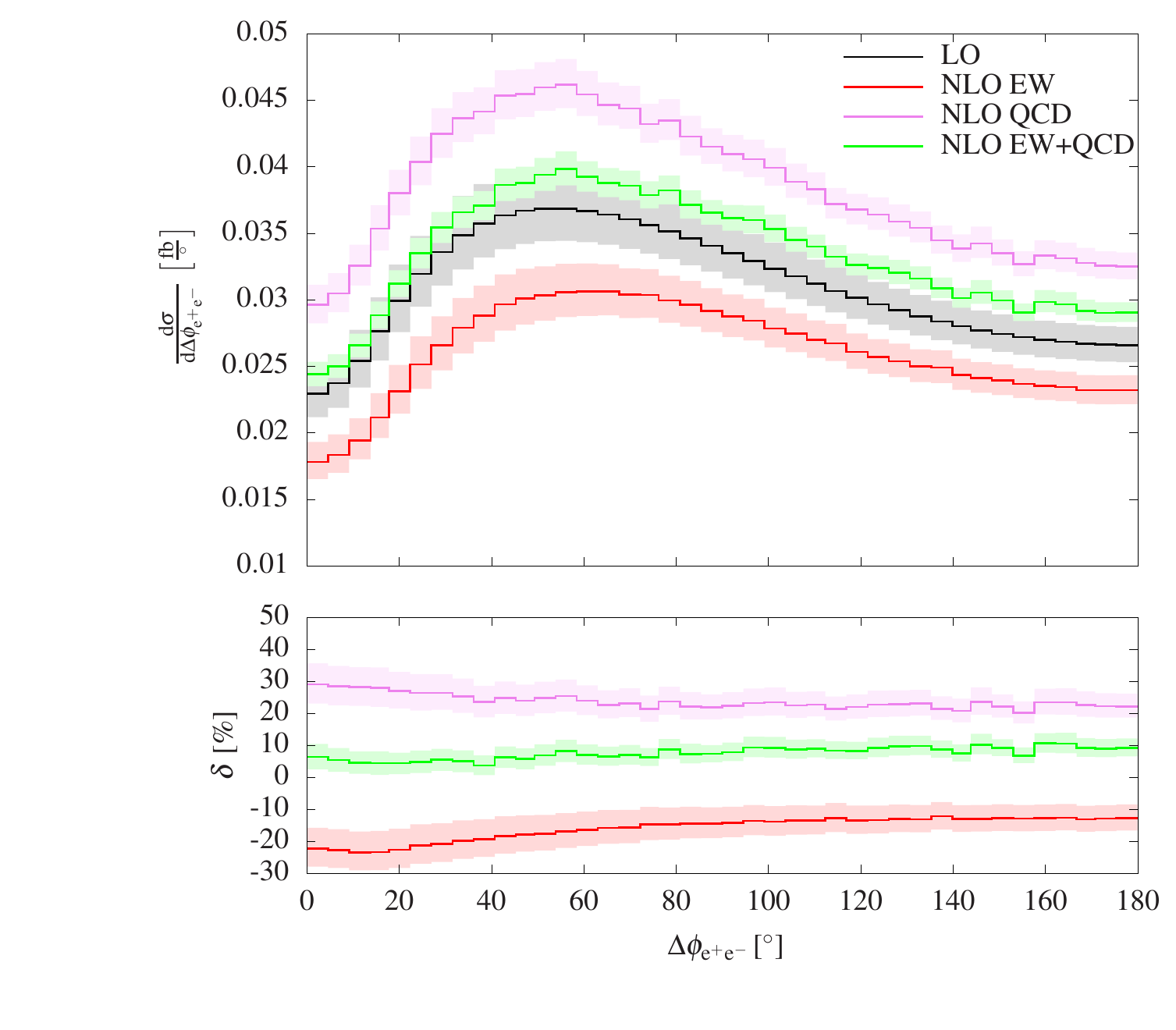}
\label{fig:aziee} 
\end{subfigure}
\begin{subfigure}{0.49\textwidth}
\centering
\subcaption{}
\includegraphics[width=1.\linewidth]{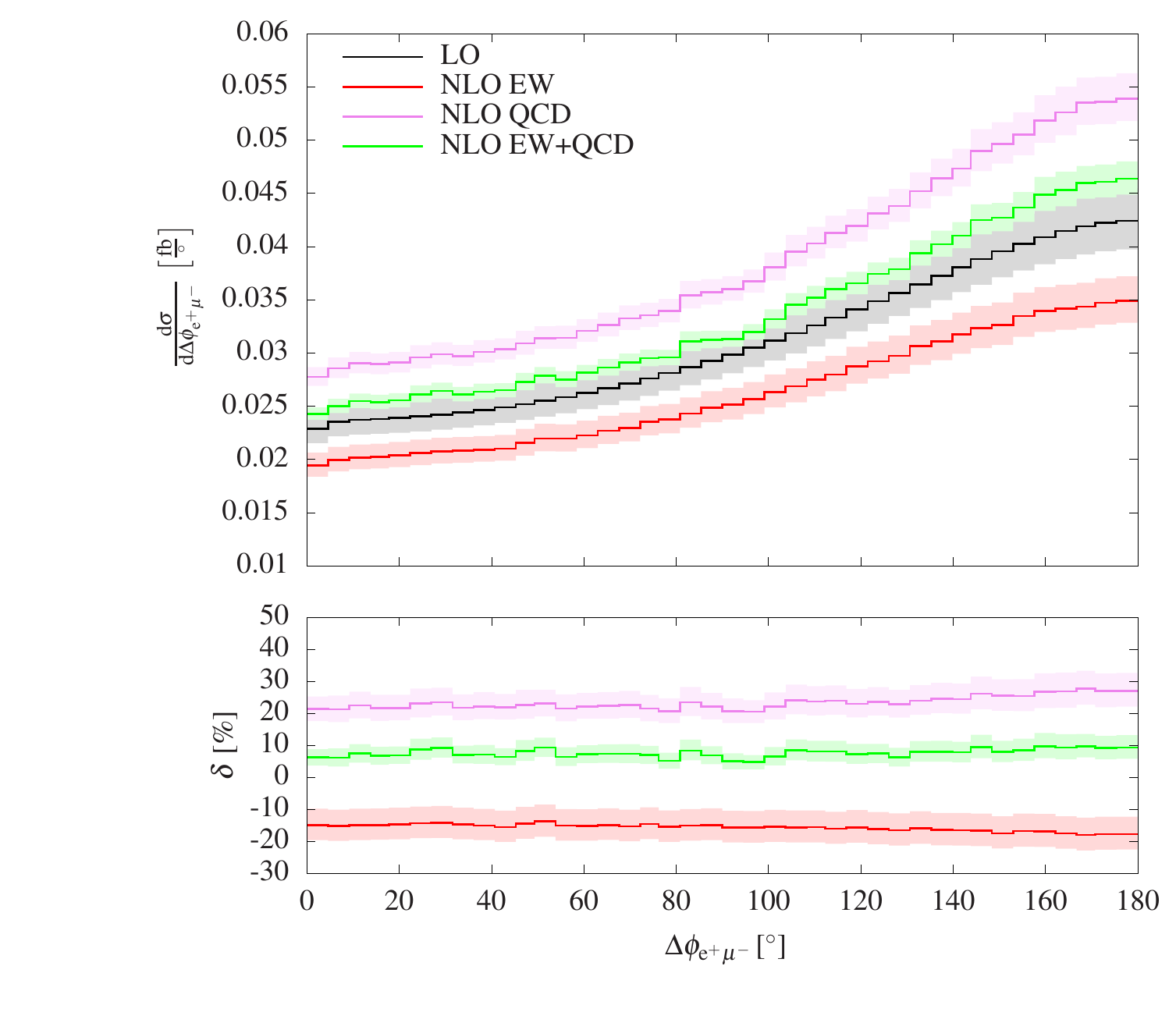}
\label{fig:aziem}
\end{subfigure}%
\par\bigskip
\begin{subfigure}{0.49\textwidth}
\centering
\subcaption{}
\includegraphics[width=1.\linewidth]{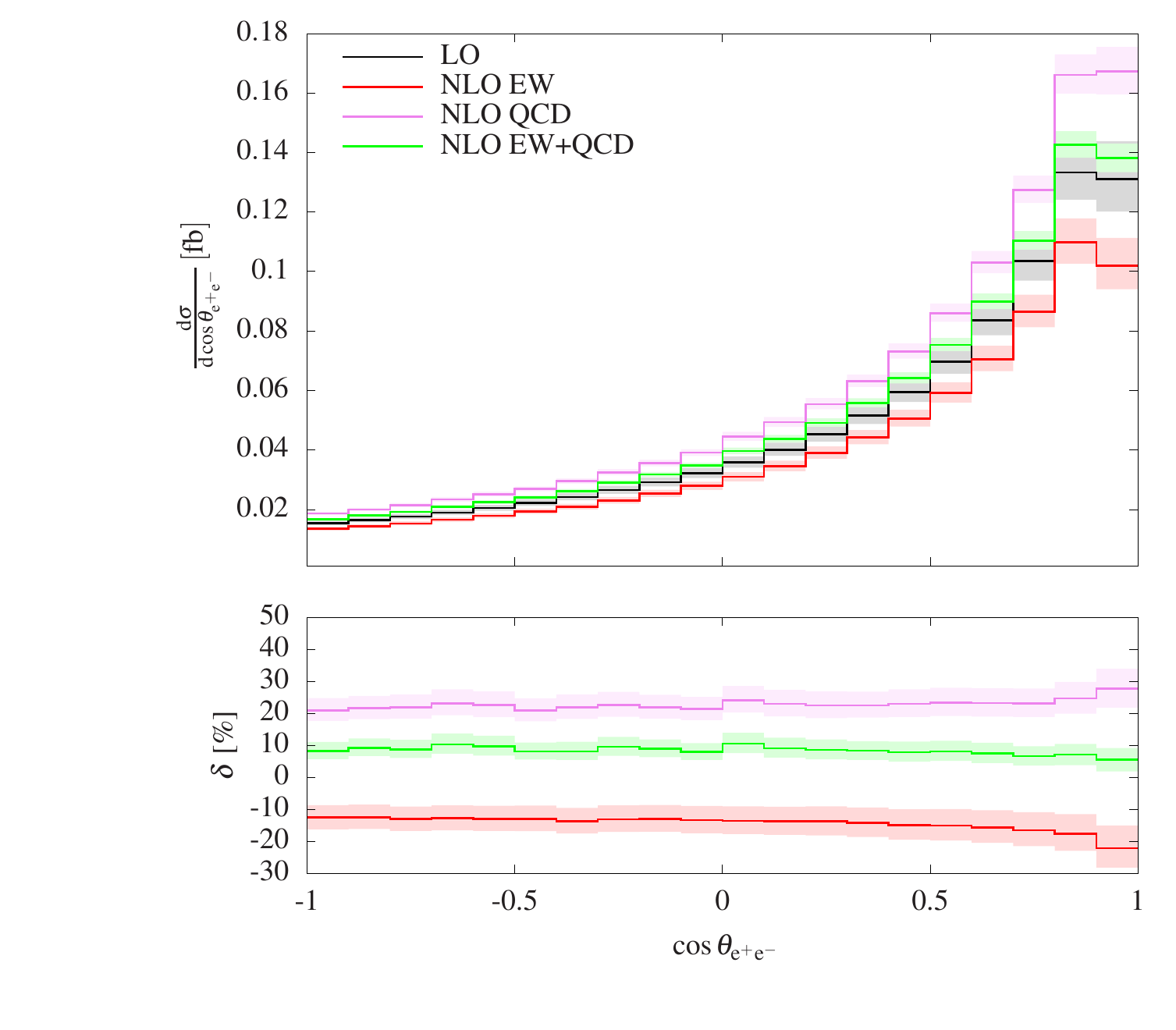}
\label{fig:cosee} 
\end{subfigure}
\begin{subfigure}{0.49\textwidth}
\centering
\subcaption{}
\includegraphics[width=1.\linewidth]{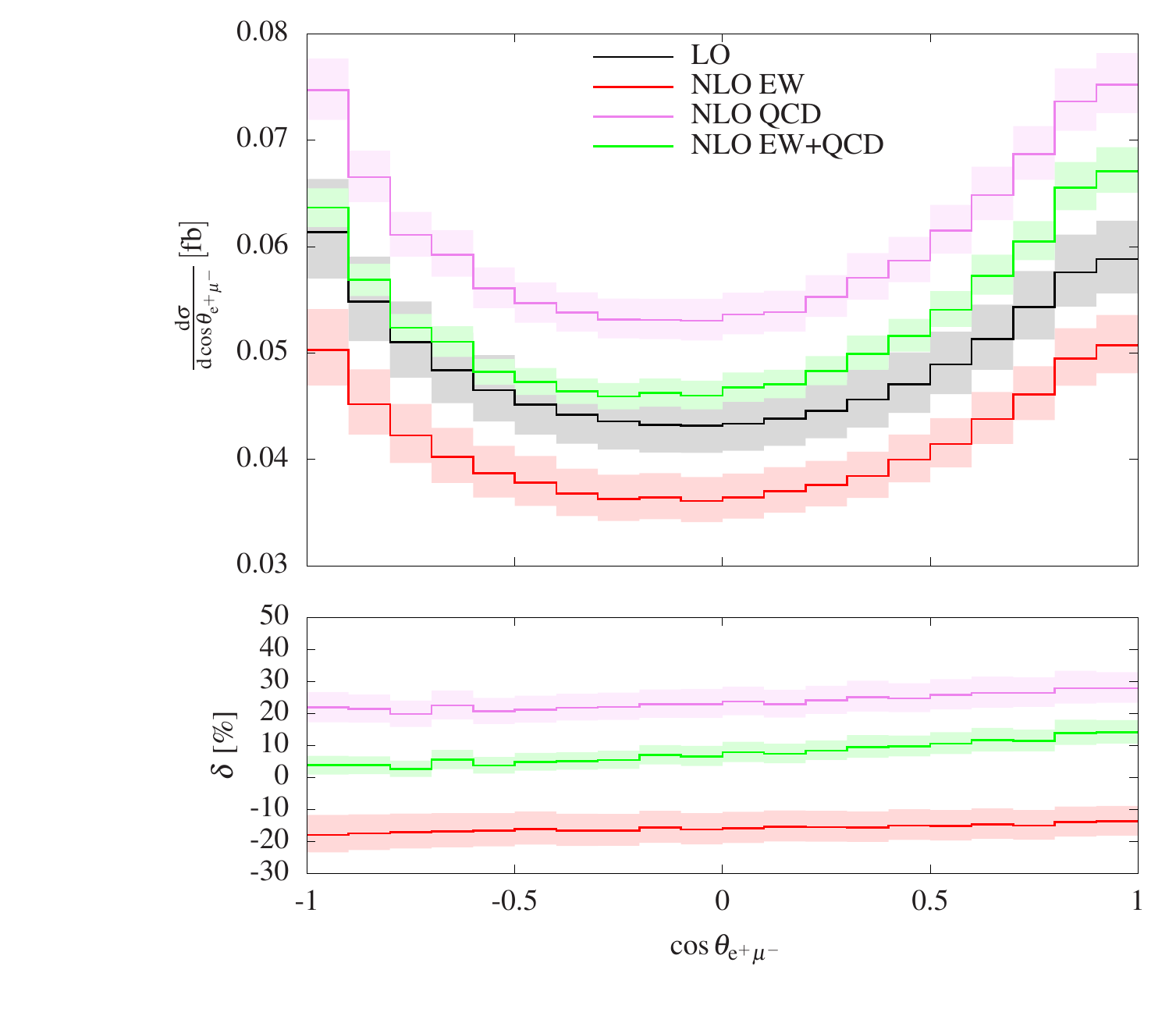}
\label{fig:cosem}
\end{subfigure}%
\vspace*{-3ex}
\caption{LO and NLO differential distributions at orders $\order{\alpha^6 }$ (LO), 
$\order{\alpha^7 }$ (NLO EW), $\order{\alphas\alpha^6}$ (NLO QCD), and NLO EW+QCD.
The upper panels show absolute predictions while the lower ones show
each contribution relative to the LO predictions. 
The observables read as follows:
azimuthal angle between the electron and the positron (top left),
azimuthal angle between the positron and the muon (top right),
cosine of the angle between the electron and the positron (bottom left), and
cosine of the angle between the positron and the muon (bottom right).}
\label{fig:NLOll}
\end{figure}

Finally, we show distributions in angular variables related to pairs
of leptons in \reffi{fig:NLOll}.
Such observables are particularly useful in $\PZ\PZ$ scattering as
they give insight in details of the scattering process which is not
possible for other VBS signatures due to the presence of neutrino(s).
We start with distributions in azimuthal-angle differences for the
electron--positron system (\reffi{fig:aziee}) and the positron--muon
system (\reffi{fig:aziem}).  Given the flavour, the first observable
relates two opposite-sign leptons originating from the same Z-boson
decay, while the second one links leptons of two different Z-boson
decays.  The shapes of the two distributions are therefore rather
different.  The distribution for the electron--positron system peaks
near $50^\circ$, while the one for the positron--muon case is maximal at
$180^\circ$.  The corrections behave also rather differently.  In the
electron--positron case, the QCD (EW) corrections are maximally
positive (negative) at low angle, leading to an overall correction
which is rather stable around $+10\%$ over the whole spectrum.  For
the positron--muon azimuthal separation, the QCD corrections slightly
increase by almost ten percent from low to large angles, while the EW
ones have the opposite trend. Thus, the overall corrections are again
rather steady at about $10\%$.  We emphasise that such cancellations
are, a priori, largely accidental and should not be taken for granted
in other VBS signatures or setups.  The last two distributions concern
the cosine of the angle between the two leptons for the same two
leptonic systems (\reffis{fig:cosee} and \ref{fig:cosem}).  It is
interesting to notice that for these distributions, again both types
of corrections do not induce large shape distortions (less than
$10\%$).  This makes such observables particularly attractive for
correlation analysis or to test models of new physics.

\FloatBarrier

\section{Conclusion}
\label{sec:conclusion}

In this article we have presented a calculation of the NLO EW and QCD
corrections of orders $\order{\alpha^7}$ and
$\order{\alphas\alpha^6}$, respectively, for the process
$\Pp\Pp\to\Pe^{+}\Pe^{-}\mu^{+}\mu^{-}\Pj\Pj+X$.  Our results include,
in particular, the NLO EW and QCD corrections to the LO EW
contribution of order $\order{\alpha^6}$, which is dominated by
vector-boson scattering (VBS) into a pair of Z~bosons.  As the full
matrix elements are used at the corresponding orders, our computation
accounts for all off-shell, non-resonant, and interference effects.
In particular, triple-boson-production processes are part
of the EW contributions.  We have included all partonic channels apart
from those involving initial-state bottom quarks or photons, which are
suppressed.

The EW corrections of order $\order{\alpha^7}$ are found to be
relatively large, in agreement with similar results obtained
previously for $\PW^\pm\PW^\pm$ and $\PW\PZ$ scattering and the
expectation that large EW corrections are an intrinsic feature of VBS
at the LHC. For the chosen fiducial cross section, the corrections are
$-16\%$ and can be well reproduced by a simple logarithmic
approximation. In the high-energy tails of distributions the EW
corrections can reach $-40\%$.  The QCD corrections of order
$\order{\alphas\alpha^6}$ exceed $20\%$ for the fiducial cross
section.  While such a magnitude for QCD corrections at the LHC is
perfectly normal, it is rather large for VBS processes.  In
particular, previous computations relying on the VBS approximation
have found QCD corrections at the percent level, in strong contrast to
our findings.  These differences are due to the fact that our
computation is done for a rather inclusive phase-space region
($M_{\Pj_1 \Pj_2} > 100\GeV$ as opposed to $M_{\Pj_1 \Pj_2} \gsim
500\GeV$ and/or $\Delta y_{\Pj_1 \Pj_2}\gsim3$ usually) and
includes tri-boson contributions.  Hence, when including real QCD
radiation at NLO, Z or W bosons decaying hadronically can become
resonant thus leading to very large corrections that are not captured
by scale variation.  We emphasise that there is nothing wrong in using
such inclusive fiducial regions as long as the theoretical simulations
used include tri-boson contributions at NLO QCD.  In particular, Monte
Carlo programs relying on the VBS approximation should not be used to
extrapolate measurements from the inclusive region to more
VBS-enriched regions.  Nonetheless, they still constitute reliable
theoretical predictions in the typical VBS regions (with
$M_{\Pj_1\Pj_2} \gsim 500\GeV$ and/or $\Delta y_{\Pj_1
    \Pj_2}\gsim3$), where the NLO QCD corrections are of 
the usual size.  The present work also shows that the proper inclusion
of tri-boson contributions is of great relevance.  We strongly
advocate for simpler and more physical measurements where tri-boson
contributions are not subtracted from the signal.  This has the
advantage to be clearly gauge invariant and not to rely on conventions
and theory predictions.  Finally, as argued in previous work,
comparing predictions including all QCD, interference, and EW
contributions with the measurements is the cleanest way of testing the
SM in such processes.

We have also analysed the composition of the LO process by comparing
contributions at orders: $\order{\alpha^6 }$ (EW
contribution), $\order{\alphas \alpha^5 }$ (interference), and
$\order{\alphas^2 \alpha^4 }$ (QCD contribution).  The findings are
in line with known observations that VBS ZZ is a challenging
channel due to the very large irreducible QCD background.  In
particular, in the fiducial region chosen, the LO EW contributions amount
to less than $10\%$.  In the LO analysis, we have also further
included the loop-induced contribution with gluon--gluon initial state
at order $\order{\alphas^4 \alpha^4}$.  It has been found to be of the
order of $10\%$ in the fiducial region, but  becomes relatively negligible
in the high-energy limit of differential distributions.
Imposing an extra VBS cut $M_{\Pj_1 \Pj_2} > 500\GeV$ enhances the EW
contribution to more than $30\%$.

We would like to point out that this calculation constitutes a further
leap in complexity with respect to previous calculations for VBS
processes.  The fact that eight charged external particles occur
increases significantly the complexity of the virtual and, in
particular, of the real corrections.  The number of partonic channels
is amplified by a factor of 5 with respect to like-sign W scattering
and 1.5 relative to WZ scattering.
As a consequence, the required CPU time is increased and efficient
book-keeping, automation and parallelisation are crucial.

Finally, we hope that these results will be useful in the current
and upcoming measurements of ZZ VBS at the LHC.  In particular, we
believe that this article contains valuable information for the
experimental collaborations when conducting their analysis.

\section*{Acknowledgements}

We are grateful to Jean-Nicolas Lang and Sandro Uccirati for
continuously supporting and improving \recola and to Robert Feger for
help with \mocanlo.  MP thanks Claude
Charlot for useful discussions.  AD, RF, and TS acknowledge financial
support by the German Federal Ministry for Education and Research
(BMBF) under contract no.~05H18WWCA1 and the German Research
Foundation (DFG) under reference numbers DE 623/6-1 and DE 623/6-2.
The research of MP has received funding from the European Research
Council (ERC) under the European Union's Horizon 2020 Research and
Innovation Programme (grant agreement no. 683211).  This work received
support from STSM Grants of the COST Action CA16108.

\bibliographystyle{JHEPmod}
\bibliography{vbs_zz}

\end{document}